\documentclass[10pt]{article} 

\usepackage[utf8]{inputenc} 

\usepackage{geometry} 
\usepackage{graphicx} 

\usepackage{booktabs} 
\usepackage{array} 
\usepackage{paralist} 
\usepackage{verbatim} 
\usepackage{subfigure}
\usepackage{natbib}
\usepackage{amssymb}
\usepackage{amsmath}
\usepackage{color}
\usepackage{rotate}
\usepackage{algorithm}
\usepackage{algpseudocode}
\usepackage{wrapfig}

\usepackage{fancyhdr} 
\usepackage{sectsty}
\allsectionsfont{\sffamily\mdseries\upshape} 

\usepackage[nottoc,notlof,notlot]{tocbibind} 
\usepackage[titles,subfigure]{tocloft} 

\title{A Numerical Approach to Solving Nonlinear Differential Equations on a Grid with Potential Applicability to Computational Fluid Dynamics}
\author{Jesper Tveit\footnote{email: jtv001@uib.no} \vspace{2mm} \\ \em{Department of Physics and Technology, University of Bergen, Norway} \\  \em{BKK Production, Kokstad, Norway}}
\date{\today}

\begin{document}

\maketitle

\begin{abstract}
A finite element method for solving nonlinear differential equations on a grid, with potential applicability to computational fluid dynamics (CFD), is developed and tested. 

The current method facilitates the computation of solutions of a high polynomial degree on a grid. A high polynomial degree is achieved by interpolating both the value, and the value of the derivatives up to a given order, of continuously distributed unknown variables.

The two-dimensional lid-driven cavity, a common benchmark problem for CFD methods, is used as a test case. It is shown that increasing the polynomial degree has some advantages, compared to increasing the number of grid-points, when solving the given benchmark problem using the current method. The current method yields results which agree well with previously published results for this test case. 


\end{abstract}

\section{Introduction}
Through development and testing in the well known case of lid-driven cavity flow (see Figure \ref{cavityConfig} for details) the current method is shown to have potential applicability to CFD. Steady state solutions of the Navier-Stokes equations for incompressible two-dimensional flow are computed for this test case with Reynolds number $\mathrm{Re} \leq 4\times 10^4$.

Obtaining a steady-state flow solution in a two-dimensional lid-driven cavity becomes increasingly challenging as the Reynolds number increases. Computing a steady-state flow solution is therefore useful as a bench-mark for the quality of numerical schemes, although the solution does not necessarily describe a physical fluid. In order to compute a solution, which accurately represents details at high Reynolds numbers, the most successful approaches have been using very high grid resolutions \citep{Erturk05,Wahba12}. The current approach obtains comparable results with much lower grid resolutions.

The current solutions have up to $9$'th order (polynomial degree) of spatial accuracy and the highest grid resolution is $135$ by $135$ grid points. Several other high order solutions to the lid-driven cavity have previously been presented. \citet{Barragy96} present solutions for the lid-driven cavity up to $\mathrm{Re} = 12500$ (as well as an under-resolved solution for $\mathrm{Re} = 16500$) with spatial accuracies from 6'th to 8'th order. Other works which present high order solutions include \citet{Schreiber82} (8'th order), and \citet{Nishida92} (10'th order). 

\citet{Wahba12} present reliable steady state solutions of comparably high Reynolds numbers ($\mathrm{Re} \leq 35 \times 10^3$) but this is the first time reliable high order (above fourth, in polynomial degree) steady state solutions for Reynolds number, $\mathrm{Re} \ge 20000$, to the lid-driven cavity in two dimensions have been presented.

The explicit definition of derivatives employed by the current method is a feature which is partially shared by the CIP method \citep{Takewaki84}, since the CIP method includes the gradient of unknown quantities as a free parameter. The CIP method is a third order method used successfully, for example, to simulate acoustic wave propagation.

The current method is a finite element type of approach and solves nonlinear differential equations through several steps. First a discretization is defined to contain information about the unknown functions in a given set of differential equations. This information includes both the value, and the value of the derivatives, of the unknown functions at specific positions (grid points) in a computational domain. Next, the differential equations are formulated as a nonlinear system of equations (weak form) depending on the information contained in the grid. This system of equations is solved through an iteration, which minimizes the square of a uniformly weighted residual. Each iteration has both a linear and a nonlinear stage, and finds an approximate solution that improves the previous approximate solution.

The specific details involved in each of these steps will be thoroughly explained in Sections \ref{framework} - \ref{itsolve}. Results of particular interest will be presented in Section \ref{results}. The complete data of all the computed results can be obtained from the author upon request.

\section{Notation and Mathematical Framework}
\label{framework}
\subsection{Notation}
\label{notation}
Square brackets will be used to identify components in matrices. A component in a two-dimensional matrix, $\mathbf{A}$, will thus be referred to as $\mathbf{A}[r,c]$, where $r$ is the row index and $c$ is the column index. Indices in an $R \times C$ matrix are defined to go from $0$ to $R-1$ (rows) and  $0$ to $C-1$ (columns). If $C = 1$, then the matrix may be referred to as a column vector and if $R = 1$, then the matrix may be referred to as a row-vector. A \textit{matrix of matrices} will be equivalent to a four dimensional matrix, $\mathbf{Q}$, where $\mathbf{Q}[r,c][\tau,\nu] \equiv \mathbf{Q}[r,c,\tau,\nu]$. If a matrix, $\mathbf{A}$, is square and nonsingular, its inverse will be written as $\mathbf{A}^{-1}$.

For brevity, the evaluation of a derivative of a function, $\Phi(\zeta)$, at a certain point, $\zeta_0$, will in unambiguous cases be written as shown on the right-hand side of Eq.(\ref{simp}):
\begin{equation}
\label{simp}
\left. \frac{\partial \Phi(\zeta) }{\partial \zeta } \right\vert_{\zeta = \zeta_0} \equiv \frac{\partial \Phi(\zeta_0) }{\partial \zeta }
\end{equation}
For functions of two variables, the first and the second argument will be referred to as the $x$-, and the $y$- component (or variable), respectively.

Mapping of indices from double index form to single index form will, unless otherwise stated, be on the form $a = b + c D$, where $D$ is a positive integer and $b,c \in \{0,\dots,D-1\}$ and $a \in \{0,\dots,D^2-1\}$. The indices, $b$ and $c$, may also a single index form of other index tuples, in which case the mapping will recursively follow the given form.

\subsection{Order of Continuity}
\label{ordCont}
Consider a discretization of a function, $f(x,y)$, on a uniform two-dimensional grid. Let the matrices $\mathbf{x}$ and $\mathbf{y}$ be composed of the $x$ and $y$ components, respectively, of the position of the grid points. Let the value of the function, $f(x,y)$, and its derivatives up to, and including, the $(\Omega-1)$'th order in each direction be explicitly defined for each grid point in terms of the (four dimensional) matrix, $\mathbf{F}$, with components given by Eq.(\ref{gridData}):
\begin{equation}
\label{gridData}
\frac{ \partial^{\alpha+\beta} f\left(\mathbf{x}\left[ k,l \right], \mathbf{y}\left[ k,l \right]\right) }{\partial x^\alpha \partial y^\beta } \equiv \mathbf{F}\left[ k,l,\alpha,\beta \right]
\end{equation}
where $\alpha \in \{0,\dots,\Omega-1 \}$, $\beta \in \{0,\dots,\Omega-1 \}$, and the indices, $k$ and $l$, identify the grid point. The discretization is then by definition continuous and has continuous derivatives up to $(\Omega-1)$'th order at the grid points $\left(\mathbf{x}\left[ k,l \right], \mathbf{y}\left[ k,l \right]\right)$. This is referred to as $\mathcal{C}^{\Omega-1}$ continuity.

\subsection{Grid Structure}
For the sake of simplicity the grid will be oriented and scaled such that the location of each grid point is uniquely determined by its indices, $k$ and $l$, as shown in Eq.(\ref{gridStruct}):
\begin{equation}
\label{gridStruct}
\left(\mathbf{x}\left[k,l\right],\mathbf{y}\left[k,l\right] \right) \equiv \left(k,l\right)
\end{equation}
defining a uniform square grid. 
\subsection{Polynomial Basis-function Expansion}
\label{basisFuncAndDegree}
Let the functions, $b_{m,n}(x,y)$, be a set of polynomial basis functions where the value of $m$ is the polynomial degree of the first variable, $x$, and the value of $n$ is the polynomial degree of the second variable, $y$.

Let the matrix of column vectors, $\mathbf{f}$, the row vector-function, $\mathbf{b}(x,y)$, and the matrix, $\mathbf{B}$, be defined as shown in Eqs.(\ref{colvec2d}-\ref{recMat2d}), respectively:
\begin{equation}
\label{colvec2d}
\mathbf{f}\left[k,l\right]{\left[\tau,0\right]} \equiv \mathbf{F}\left[ k+i,l+j,\alpha,\beta \right]
\end{equation}
\begin{equation}
\label{rowvec2d}
\mathbf{b}{\left[0,m + Nn \right]}(x,y) \equiv b_{m,n}(x,y)
\end{equation} 
\begin{equation}
\label{recMat2d}
\mathbf{B}\left[\tau,m + Nn\right] \equiv \frac{\partial^{\alpha+\beta}b_{m,n}(i,j)}{\partial x^\alpha \partial y^\beta}
\end{equation}
where $\tau = \alpha + \beta \Omega + \Omega^2(i + 2j)$, $i \in \{ 0,1 \}$, $j \in \{ 0,1 \}$, $m \in \{ 0,\dots,N-1 \}$, $n \in \{ 0,\dots,N-1 \}$ and $N = 2\Omega$. It follows that each column vector in the matrix, $\mathbf{f}$, then has $(2\Omega)^2 = N^2$ components, that the matrix, $\mathbf{B}$, is a $N^2 \times N^2$ square matrix and that the row vector-function, $\mathbf{b}(x,y)$, has $N^2$ components. 

The four neighboring points: $\left(\mathbf{x}\left[k,l\right],\mathbf{y}\left[k,l\right] \right)$, $\left(\mathbf{x}\left[k+1,l\right],\mathbf{y}\left[k+1,l\right] \right)$, $\left(\mathbf{x}\left[k,l+1\right],\mathbf{y}\left[k,l+1\right] \right)$ and $\left(\mathbf{x}\left[k+1,l+1\right],\mathbf{y}\left[k+1,l+1\right] \right)$, surround a square region which will be referred to as the $k,l$'th \textit{grid-cell}. 

Within the $k,l$'th grid-cell, the function, $f(x,y)$, may be approximated by a weighted sum of the polynomial basis functions, $b_{m,n}(x,y)$, written in matrix form in Eq.(\ref{recon2d}):
\begin{equation}
\label{recon2d}
f(x,y) = \mathbf{b}(x',y') \mathbf{B}^{-1} \mathbf{f}\left[k,l\right] + \mathcal{O}(x'^N + y'^N)  \approx
\mathbf{b}(x',y') \mathbf{B}^{-1} \mathbf{f}\left[k,l\right]
\end{equation} 
where $x' = x - \mathbf{x}\left[k,l\right]$ and $y' = y - \mathbf{y}\left[k,l\right]$. 

From the definitions, Eqs.(\ref{colvec2d}-\ref{recMat2d}), it is clear that the approximation, $\mathbf{b}(x',y') \mathbf{B}^{-1} \mathbf{f}\left[k,l\right]$, matches up exactly with the discretization, $\mathbf{F}$, at the four grid points,  i.e.: 
\begin{equation}
\label{show}
\mathbf{F}\left[ k + i,l + j,\alpha,\beta \right] = \frac{\partial^{\alpha+\beta} \mathbf{b}(i,j) }{\partial x^\alpha \partial y^\beta}
\mathbf{B}^{-1} \mathbf{f}\left[k,l\right]
\end{equation}
where $i \in \{0,1\}$ and $j \in \{0,1\}$ as previously.

\subsection{Hermite Splines}
The idea of approximating a function by sampling both its value and the value of its derivatives is known as Hermite interpolation. The approximation, $ \mathbf{b}(x',y') \mathbf{B}^{-1} \mathbf{f}\left[k,l\right]$, of the function, $f(x,y)$, is a two-dimensional generalization of a Hermite spline, equivalent to recursively interpolating a set of Hermite splines.

\subsection{Choice of Basis Functions}
\label{basisChoice}
The \textit{condition number}, $\mathrm{cond}(\mathbf{B})$, defined in Eq.(\ref{condDef}), gives an estimate of the relative numerical accuracy of the matrix product, $\mathbf{B}^{-1} \mathbf{f}\left[k,l\right]$.
\begin{equation}
\label{condDef}
\mathrm{cond}\left(\mathbf{B}\right) \equiv \frac{\sigma_{\mathrm{max}}\left(\mathbf{B}\right)}{\sigma_{\mathrm{min}}\left(\mathbf{B}\right)}
\end{equation}
In Eq.(\ref{condDef}), $\sigma_{\mathrm{max}}\left(\mathbf{B}\right)$ is the largest singular value of $\mathbf{B}$, and $\sigma_{\mathrm{min}}\left(\mathbf{B}\right)$ is the smallest singular value of $\mathbf{B}$. When numerically solving a linear system, $\mathbf{f} = \mathbf{B} \mathbf{c}$, using floating point numbers with machine precision, $\epsilon_{m}$, an error of order $\mathcal{O}\left(\epsilon_{m} \mathrm{cond}\left( \mathbf{\textbf{B}} \right)\right)$ should be expected. The reader may refer to \citet[pg. 95]{trefethen1997} for a more detailed explanation of the condition number and numerical accuracy of linear equation systems.

The condition number, $\mathrm{cond}(\mathbf{B})$, depends on the choice of basis functions, $b_{m,n}(x,y)$, and on the order of continuity (in other words, the value of $\Omega$). 

For the computations in this paper, the basis functions, $b_{m,n}(x,y)$, are defined in terms of the Bernstein polynomials, $\mathcal{B}_{\lambda,\Lambda}(x)$, given in Eq.(\ref{bernstein1d}):
\begin{equation}
\label{bernstein1d}
\mathcal{B}_{\lambda,\Lambda}(x) = \binom{\Lambda}{\lambda} x^\lambda (1-x)^{\Lambda-\lambda}, \;  \lambda \in \{ 0, \dots, \Lambda \}
\end{equation}
as
\begin{equation}
\label{bernstein2d}
b_{m,n}(x,y) \equiv \mathcal{B}_{m,\Omega-1}(x) \mathcal{B}_{n,\Omega-1}(y)
\end{equation}
Table \ref{condnums} shows that the condition number of the matrix, $\mathbf{B}$, increases exponentially with the value of $\Omega$. The machine precision is a limiting factor for the computation of the function approximation, Eq.(\ref{recon2d}). For higher orders of continuity, it may be considered an ill-conditioned system. As a result one should not expect the coefficients of the function approximation, Eq.(\ref{recon2d}), to be accurate down to machine epsilon. For this reason, the main results presented in this paper have been computed using 64 bit floating point numbers (\textit{double} in C++ syntax) rather than the more common 32 bit float. The corresponding ISO C standard definition of machine epsilon is $2^{-52} \approx 2.2 \times 10^{-16}$, referred to in this paper as $\epsilon_{64}$.

\begin{table}[h]
\centering
\begin{tabular}{ccc}
\toprule
      & $\epsilon_{64} \mathrm{cond}\left(\mathbf{B}\right)$ & $\epsilon_{64} \mathrm{cond}\left(\mathbf{B}\right)$ \\
  $\Omega$ & with $b_{m,n}(x,y)$ & with $x^m y^n$  \\  
   2 & $8.5 \times 10^{-15}$ & $1.3 \times 10^{-13}$ \\
   3 & $1.7 \times 10^{-12}$ & $1.3 \times 10^{-10}$ \\
   4 & $8.2 \times 10^{-10}$ & $5.0 \times 10^{-7}$ \\
   5 & $7.5 \times 10^{-7}$ & $4.4 \times 10^{-3}$ \\
   6 & $1.1 \times 10^{-3}$ & $23.5$ \\ \bottomrule
\end{tabular}
\caption{Estimated floating point errors for different values of $\Omega$. The matrix, $\mathbf{B}$, is defined in Eq.(\ref{recMat2d}) and the condition number, $\mathrm{cond}\left(\mathbf{B}\right)$, is defined by Eq.(\ref{condDef}). This is the estimated precision of the numerical computation of the quantity $\mathbf{B}^{-1} \mathbf{f}$ using 64 bit floating point numbers with machine precision, $\epsilon_{64} \approx 2.2 \times 10^{-16}$, (ISO C standard). The first column shows different values of $\Omega$, corresponding to $\mathcal{C}^{\Omega-1}$ continuity. The second column shows, $\epsilon_{64} \mathrm{cond}\left(\mathbf{B}\right)$, constructed with the basis functions, $b_{m,n}(x,y)$, as given by Eq.(\ref{bernstein2d}). The third column shows what the expected  precision would be if the monomial basis functions, $x^my^n$, of equal degree were used to construct the matrix $\mathbf{B}$ instead of $b_{m,n}(x,y)$.}
\label{condnums}
\end{table}

\section{Discretization the Navier--Stokes Equations}
\subsection{Navier--Stokes Equations for Steady State Incompressible Flow in Two Dimensions}
\label{navSto2d}
The grid has $L \times L$ grid-points at positions defined in Eq.(\ref{gridStruct}). The indices, $k$ and $l$, then have values ranging from $0$ to $L-1$ and the grid is square with length and width equal to $L-1$. The computational domain is thus the two dimensional interval $[0,L-1] \times [0,L-1]$ (see Figure \ref{cavityConfig}).

The Navier--Stokes Equations for steady state incompressible flow in two dimensions, where the physical variables have been scaled with appropriate scales, read
\begin{subequations}
\begin{align}
\label{momx}
0 &= u \frac{\partial u }{\partial x } + v \frac{\partial u}{\partial y } + \frac{\partial p}{ \partial x} - \frac{1}{\mathrm{Re}'} \left( \frac{\partial^2 u}{\partial x^2} + \frac{\partial^2 u}{\partial y^2} \right) \\
\label{momy}
0 &= u \frac{\partial v}{\partial x } + v \frac{\partial v}{\partial y } + \frac{\partial p}{ \partial y} - \frac{1}{\mathrm{Re}'} \left( \frac{\partial^2 v}{\partial x^2} + \frac{\partial^2 v}{\partial y^2} \right) \\
\label{con} 
0 &= \frac{\partial u}{\partial x} + \frac{\partial v}{\partial y}
\end{align}
\end{subequations}
where $u$ and $v$ are the $x$-- and $y$--components of the flow velocity, respectively, $p$ is the pressure and $\mathrm{Re}' = \mathrm{Re}/(L-1)$. The definition of the Reynolds number, $\mathrm{Re}$, is the same as in the references \citep{Ghia82,Erturk05,Wahba12}. However, since the the computational domain used by the references is the two dimensional interval $[0,1] \times [0,1]$, the Reynolds number, $\mathrm{Re}$, must be scaled with the size of the current domain, $L-1$, so that Eqs.(\ref{momx}-\ref{con}) remain mathematically equivalent to the equations solved in the references. 

The variables $u$, $v$ and $p$ will be referred to as the flow-variables and are functions of the two variables $x$ and $y$. Additionally, the pair $(u,v)$ will be referred to as the flow-velocity. 

The pressure-velocity form of the Navier--Stokes equations (Eqs.(\ref{momx}-\ref{con})) is usually transformed into an equivalent \textit{vorticity-streamfunction} form which, in the two dimensional case, has one less flow variable to deal with. Most published papers dealing with the lid-driven cavity in two dimensions use the vorticity-streamfunction form. In the current work, however, we solve for the pressure and velocity directly.

\subsection{Boundary Conditions}
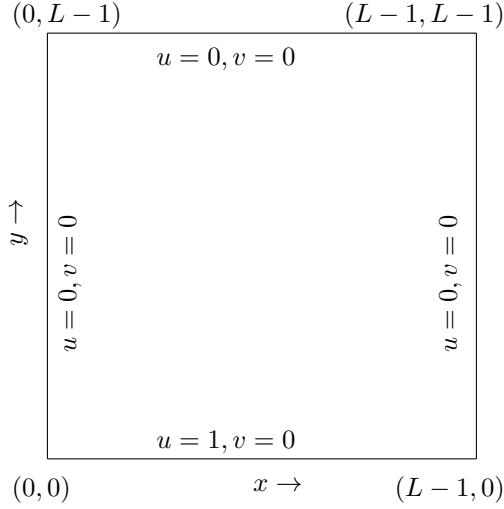
\begin{figure}[h]
\centering
\setlength{\unitlength}{0.48\textwidth}
\begin{minipage}[c]{\unitlength}
\begin{picture}(1,1)
\put(0.1,0.1){ \line(1,0){0.8} }
\put(0.1,0.1){ \line(0,1){0.8} }
\put(0.1,0.9){ \line(1,0){0.8} }
\put(0.9,0.1){ \line(0,1){0.8} }
\put(0.05,0.03){\color[rgb]{0,0,0} \makebox(0,0)[lb]{\smash{$\left(0,0\right)$}}}
\put(0.97,0.03){\color[rgb]{0,0,0} \makebox(0,0)[rb]{\smash{$\left(L-1,0\right)$}}}
\put(0.05,0.92){\color[rgb]{0,0,0} \makebox(0,0)[lb]{\smash{$\left(0,L-1\right)$}}}
\put(0.97,0.92){\color[rgb]{0,0,0} \makebox(0,0)[rb]{\smash{$\left(L-1,L-1\right)$}}}
\put(0.07,0.50){\color[rgb]{0,0,0}\rotatebox{90}{\makebox(0,0)[lb]{\smash{$y \rightarrow$}}}}
\put(0.50,0.04){\color[rgb]{0,0,0} \makebox(0,0)[lb]{\smash{$x \rightarrow$}}}
\put(0.32,0.12){\color[rgb]{0,0,0} \makebox(0,0)[lb]{\smash{$u = 1, v = 0$}}}
\put(0.32,0.84){\color[rgb]{0,0,0} \makebox(0,0)[lb]{\smash{$u = 0, v = 0$}}}
\put(0.17,0.30){\color[rgb]{0,0,0}\rotatebox{90}{\makebox(0,0)[lb]{\smash{$u = 0, v = 0$}}}}
\put(0.88,0.30){\color[rgb]{0,0,0}\rotatebox{90}{\makebox(0,0)[lb]{\smash{$u = 0, v = 0$}}}}
\end{picture}
\end{minipage} \hspace{0.05\unitlength}
\begin{minipage}[c]{0.7\unitlength}
\caption{This figure shows the computational domain of the grid and its boundary conditions. The flow velocity, $u(x,y)$ and $v(x,y)$, along the edges is given. The lower edge, where $y = 0$, is a "lid" which drives the flow by sliding horizontally (positive $x$- direction) at a constant speed equal to one. At the three remaining edges, both components of the flow velocity are zero. This system is referred to as a lid-driven cavity.}
\label{cavityConfig}
\end{minipage}
\end{figure}

The no-slip Dirichlet boundary condition is imposed on the flow velocity, $\left(u(x,y),v(x,y)\right)$. Figure \ref{cavityConfig} shows the details of the boundary of the computational domain of the grid. This test case is known as the lid-driven cavity for two dimensions. 
No boundary values are required for the pressure during the iterative solution process. The origin of the pressure is implicitly determined by the initial condition for the iteration (see Section \ref{itsolve}).

\subsection{Navier-Stokes Equations in Matrix Form}
\label{weakform}
The flow-variables, velocity and pressure, are discretized as shown in Section \ref{framework}, with $\mathbf{U}$,$\mathbf{V}$, and $\mathbf{P}$ being the discrete counterparts of $u(x,y)$, $v(x,y)$ and $p(x,y)$, respectively. In each grid-cell, the flow variables are approximated by
\begin{subequations}
\begin{align}
u(x,y) &= \mathbf{b}(x',y') \mathbf{B}^{-1} \mathbf{u}\left[k,l\right] + \mathcal{O}(x'^N + y'^N) \\
v(x,y) &= \mathbf{b}(x',y') \mathbf{B}^{-1} \mathbf{v}\left[k,l\right] + \mathcal{O}(x'^N + y'^N) \\
p(x,y) &= \mathbf{b}(x',y') \mathbf{B}^{-1} \mathbf{p}\left[k,l\right] + \mathcal{O}(x'^N + y'^N)
\end{align}
\end{subequations}
where $\mathbf{u}$, $\mathbf{v}$ and $\mathbf{p}$ are defined in terms of $\mathbf{U}$,$\mathbf{V}$, and $\mathbf{P}$ in the same way as $\mathbf{f}$ was defined in terms of $\mathbf{F}$ in Section \ref{framework}.

Let the row vector-functions, $\mathbf{c}_{\alpha,\beta}(x,y)$ and $\mathbf{s}(x,y)$, and the matrix of row vector-functions, $\mathbf{m}(x,y)$, be defined as shown in Eq.(\ref{diffDiscC}), Eq.(\ref{diffDiscS}) and Eq.(\ref{diffDiscM}):
\begin{equation}
\label{diffDiscC}
\mathbf{c}_{\alpha,\beta}(x,y) \equiv \frac{\partial^{\alpha + \beta} \mathbf{b}(x,y)}{\partial x^\alpha \partial y^\beta} \mathbf{B}^{-1}
\end{equation}
\begin{equation}
\label{diffDiscS}
\mathbf{s}(x,y) \equiv - \frac{L-1}{\mathrm{Re}} \left(\mathbf{c}_{2,0}(x,y) + \mathbf{c}_{0,2}(x,y) \right)
\end{equation}
\begin{equation}
\label{diffDiscM}
\mathbf{m}\left[k,l\right](x,y) \equiv \left( \mathbf{c}_{0,0}(x,y) \mathbf{u}[k,l] \right) \mathbf{c}_{1,0}(x,y) 
                  + \left( \mathbf{c}_{0,0}(x,y) \mathbf{v}[k,l] \right) \mathbf{c}_{0,1}(x,y) + \mathbf{s}(x,y) 
\end{equation}
The Navier--Stokes equations, Eqs.(\ref{momx}-\ref{con}), are then approximated, within a grid-cell as
\begin{subequations}
\begin{align}
\label{ddmx}
0 =& \mathbf{m}[k,l] \mathbf{u}[k,l] + \mathbf{c}_{1,0} \mathbf{p}[k,l] + \mathcal{O}(x^{N-2} + y^{N-2}) \\
\label{ddmy}
0 =& \mathbf{m}[k,l] \mathbf{v}[k,l] + \mathbf{c}_{0,1} \mathbf{p}[k,l] + \mathcal{O}(x^{N-2} + y^{N-2}) \\
\label{ddcon}
0 =& \mathbf{c}_{1,0} \mathbf{u}[k,l] + \mathbf{c}_{0,1} \mathbf{v}[k,l] + \mathcal{O}(x^{N-1} + y^{N-1})
\end{align}
\end{subequations}
As indicated in Eqs.(\ref{ddmx}-\ref{ddcon}), the formal polynomial order of accuracy is reduced due to the differentation with respect to $x$ and $y$. From now on, the indication of polynomial order of accuracy, $\mathcal{O}(\dots)$, will be omitted.

Eqs.(\ref{ddmx}-\ref{ddcon}), can be written in the form of a single matrix equation, as shown in Eq.(\ref{cellSys}):
\begin{equation}
\label{cellSys}
\mathbf{0} = \mathbf{E}[k,l](x,y) \mathbf{z}[k,l]
\end{equation}
where the $3 \times 3N^2$ matrix-functions, $\mathbf{E}[k,l](x,y)$, and the $3N^2 \times 1$ column vectors, $\mathbf{z}[k,l]$, are defined by Eq.(\ref{cellMat}) and Eq.(\ref{cellVec}), respectively, as:
\begin{equation}
\label{cellMat}
\mathbf{E}[k,l](x,y) \equiv  \left[ \begin{array}{ccc}
\mathbf{m}[k,l](x,y) & \mathbf{0} & \mathbf{c}_{1,0}(x,y) \\
\mathbf{0} & \mathbf{m}[k,l](x,y) & \mathbf{c}_{0,1}(x,y) \\
\mathbf{c}_{1,0}(x,y) & \mathbf{c}_{0,1}(x,y) & \mathbf{0}
\end{array} \right]
\end{equation}
and
\begin{equation}
\label{cellVec}
\mathbf{z}[k,l] \equiv \left[ \begin{array}{c}
\mathbf{u}[k,l] \\
\mathbf{v}[k,l] \\
\mathbf{p}[k,l]
\end{array} \right]
\end{equation}

\subsection{Boundary Components of the Grid}
\label{boundaryComp}
In accordance with the definition of the grid structure, Eq.(\ref{gridStruct}), and the boundary conditions for the lid-driven cavity (Figure \ref{cavityConfig}), a grid component of the flow velocity, $\mathbf{U}[k,l,\alpha,\beta]$ or $\mathbf{V}[k,l,\alpha,\beta]$, will be defined to be a constant \textit{boundary component} if $\left( l = 0 \lor l = L-1 \right) \land \beta = 0$ (boundary parallel to the $x$-axis) or if $\left( k = 0 \lor k = L-1 \right) \land \alpha = 0$ (boundary parallel to the $y$-axis). Note that \textit{derivatives, parallel to the boundary, at the boundary} are also defined as boundary components. The values of the boundary components are all zero, except in the cases given by Eq.(\ref{lidBoundary}) and Figure \ref{cornerModel}. The components which are not defined to be boundary components will be referred to as \textit{internal components} or as \textit{internal flow components}.
\begin{equation}
\label{lidBoundary}
\mathbf{U}\left[ k,0,0,0 \right] =
\begin{cases}
1, \; \mbox{ for } \; 0 < k < L-1 \\
\frac{1}{2}, \; \mbox{ for } \; k \in \{ 0, L-1 \}
\end{cases}
\end{equation}
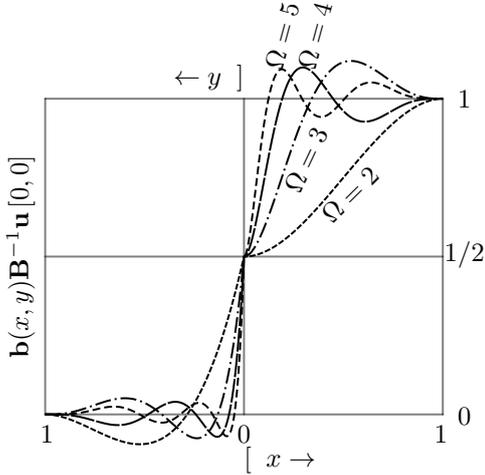
\begin{figure}[h]
\centering
\setlength{\unitlength}{0.42\textwidth}
\begin{minipage}[c]{\unitlength}
\begin{picture}(1,1.1)
    \put(0,0){\includegraphics[width=\unitlength]{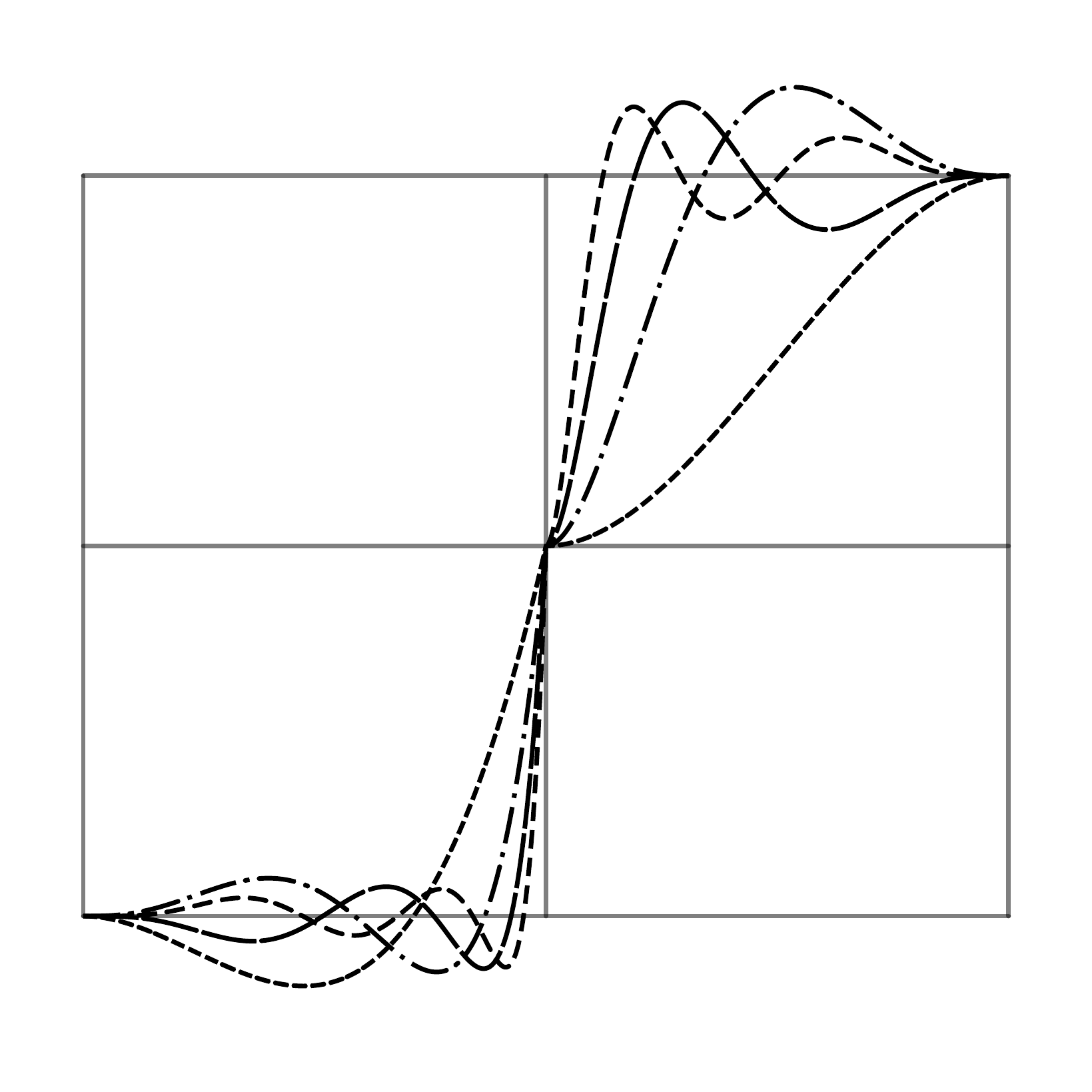}}
	\put(0.5,0.05){\color[rgb]{0,0,0} \makebox(0,0)[lb]{\smash{$ [ \; \;  x \rightarrow$}}}
    \put(0.5,0.87){\color[rgb]{0,0,0} \makebox(0,0)[rb]{\smash{$ \leftarrow y \; \; ]$}}}
	\put(0.04,0.50){\color[rgb]{0,0,0}\rotatebox{90}{\makebox(0,0)[cb]{\smash{$\mathbf{b}(x,y) \mathbf{B}^{-1} \mathbf{u}\left[0,0\right]$}}}}
	\put(0.97,0.82){\color[rgb]{0,0,0} \makebox(0,0)[cb]{\smash{$1$}}}
	\put(0.97,0.49){\color[rgb]{0,0,0} \makebox(0,0)[cb]{\smash{$1/2$}}}
	\put(0.97,0.14){\color[rgb]{0,0,0} \makebox(0,0)[cb]{\smash{$0$}}}
	\put(0.74,0.62){\color[rgb]{0,0,0}\rotatebox{45}{\makebox(0,0)[cb]{\smash{$\Omega = 2$}}}}
	\put(0.65,0.70){\color[rgb]{0,0,0}\rotatebox{65}{\makebox(0,0)[cb]{\smash{$\Omega = 3$}}}}
	\put(0.67,0.97){\color[rgb]{0,0,0}\rotatebox{75}{\makebox(0,0)[cb]{\smash{$\Omega = 4$}}}}
	\put(0.60,0.97){\color[rgb]{0,0,0}\rotatebox{75}{\makebox(0,0)[cb]{\smash{$\Omega = 5$}}}}
     \put(0.92,0.10){\color[rgb]{0,0,0} \makebox(0,0)[cb]{\smash{$1$}}}
     \put(0.08,0.10){\color[rgb]{0,0,0} \makebox(0,0)[cb]{\smash{$1$}}}
     \put(0.50,0.10){\color[rgb]{0,0,0} \makebox(0,0)[cb]{\smash{$0$}}}
\end{picture}
\end{minipage}\hspace{0.1\unitlength}
\begin{minipage}[c]{1.1\unitlength}
\caption{This figure shows how the boundary value of the $x$-component of the velocity, $u(x,y)$, is approximated near the lower left corner ($x = y = 0$) of the boundary for different values of $\Omega$. The graphs on the right hand side of the figure are for $y = 0$ out to the nearest grid point along the $x$-axis. The graphs on the left hand side of the figure are for $x = 0$ out the to the nearest grid point along the $y$-axis. By using a standard least squares algorithm (see for example \citet[pg. 437]{anton2000}), the components $\mathbf{U}[0,0,\{2,\dots,\Omega-1\},0]$ are set to produce the best fit to the unit boundary condition and the components $\mathbf{U}[0,0,0,\{1,\dots,\Omega-1\}]$ are set to produce the best fit to the zero boundary condition. Note that the derivative of $u$ with respect to $x$ at the corner, $\mathbf{U}[0,0,1,0]$, must be zero since the derivative of $v$ with respect to $y$ at the corner, $\mathbf{V}[0,0,0,1]$, is zero (otherwise the continuity equation, Eq.(\ref{con}), would not be satisfied at these points). The velocity near the lower right corner is modeled in the same way (mirrored along the $x$-axis).}
\label{cornerModel}
\end{minipage}
\end{figure}

\section{Linear Approximation of the Navier-Stokes Equations}
\subsection{Integrated Error-Squared}
\label{subCell}
In this subsection (\ref{subCell}), the grid-cell indices, $[k,l]$, will be omitted occasionally for the sake of brevity. Unless otherwise stated, the equations and definitions will be understood to correspond to a single, arbitrary grid-cell. 

The error-squared, $R^2$, for each grid-cell will be defined, using the approximated Navier-Stokes equations on matrix form (Eq.(\ref{cellSys})), as
\begin{equation}
\label{cellError}
R^2 \equiv \frac{1}{2} \int \limits_{0}^{1} \int \limits_{0}^{1} \mathbf{z}^T \mathbf{E}^T(x,y) \mathbf{E}(x,y) \mathbf{z} \,  \mathrm{d}x \mathrm{d}y
\end{equation}

A linear approximation of the derivatives of the error-squared, $R^2[k,l]$, with respect to the components of the column vector, $\mathbf{z}$, is
\begin{equation}
\label{cellErrorDiff}
\frac{\partial R^2}{\partial \mathbf{z} } \approx \left( \int \limits_{0}^{1} \int \limits_{0}^{1} \mathbf{E}^T(x,y) \mathbf{E}(x,y) \,  \mathrm{d}x \mathrm{d}y \right) \mathbf{z}
\end{equation}
The derivative, given in Eq.(\ref{cellErrorDiff}), is an approximation because the matrix, $\mathbf{E}$, is taken to be constant (while, in fact, it depends on the flow velocity through its dependence on the matrix, $\mathbf{m}[k,l]$).

It is possible to compute the integral, Eq.(\ref{cellErrorDiff}), analytically. But, by using numerical integration, the following procedure is more flexible (with future modifications and extensions in mind). 

Let the points, $(x_s,y_s)$, for $s \in \{0,\dots,S^2-1\}$ form a uniform set of sample positions, defined in Eq.(\ref{samples}):
\begin{equation}
\label{samples}
\left(x_s,y_s\right) \equiv \left( \frac{1+s_x}{1+S}, \frac{1+s_y}{1+S} \right)
\end{equation}
where $s = s_x + s_yS$ and $s_x,s_y \in \{0,\dots,S-1\}$. From Eq.(\ref{samples}) it is clear that $0 < x_s < 1$ and $0 < y_s < 1$. The approximation of the derivatives of the error-squared, $R^2[k,l]$, with respect to the components of the column vector, $\mathbf{z}[k,l]$, where the integral has been replaced by a sum over the samples, $(x_s,y_s)$, read:
\begin{equation}
\label{cellErrorDiffNum}
\frac{\partial R^2}{\partial \mathbf{z} } \approx \left(
\frac{1}{S^2} \sum_{s = 0}^{S^2-1} \mathbf{E}^T(x_s,y_s) \mathbf{E}(x_s,y_s) \right) \mathbf{z} 
\end{equation}

\subsection{Sub-cell System to Grid-wide System}
Let the $(L-1) \times (L-1)\times 3N^2 \times 3N^2$ matrix, $\mathbf{R}$, be defined by Eq.(\ref{cellSysApp}):
\begin{equation}
\label{cellSysApp}
\frac{1}{S^2} \sum_{s = 0}^{S^2-1} \mathbf{E}[k,l]^T(x_s,y_s) \mathbf{E}[k,l](x_s,y_s)  \equiv \mathbf{R}[k,l]
\end{equation}
with the matrix, $\mathbf{E}$, as defined in Eq.(\ref{cellMat}). The definition, Eq.(\ref{cellSysApp}), implies that the the $3N^2 \times 3N^2$ matrices, $\mathbf{R}[k,l]$, are symmetric. The grid-cell system, Eq.(\ref{cellErrorDiffNum}), may be written as shown in Eq.(\ref{cellSysMatApp}):
\begin{equation}
\label{cellSysMatApp}
\frac{\partial R^2[k,l]}{\partial \mathbf{z} } \approx \mathbf{R}[k,l] \mathbf{z}[k,l]
\end{equation}
To find an approximate minimum of the error-squared, $R^2[k,l]$, for all the grid-cells, we formulate the equation system given in Eq.(\ref{cellSysMatApp2}) from which a solution for the internal flow components is implied.
\begin{equation}
\label{cellSysMatApp2}
\mathbf{R}[k,l] \mathbf{z}[k,l] = \mathbf{0}
\end{equation}
Recall that the matrices, $\mathbf{u}$, $\mathbf{v}$ and $\mathbf{p}$ (and thus also $\mathbf{z}$), are defined in terms of the three four-dimensional matrices $\mathbf{U}$,$\mathbf{V}$, and $\mathbf{P}$. From the definition given in Eq.(\ref{colvec2d}) it is clear that there is an overlap between some of the components in the vectors, $\mathbf{u}[k,l]$, $\mathbf{v}[k,l]$ and  $\mathbf{p}[k,l]$, for different values of the cell indices, $k$ and $l$ (for example, the components $\mathbf{u}[k,l][\alpha + \beta \Omega + 3\Omega^2,0]$ and $\mathbf{u}[k+1,l+1][\alpha + \beta \Omega,0]$, with $\alpha,\beta \in \{0,\dots,\Omega-1\}$, correspond to the same components in the matrix, $\mathbf{U}$, and are then by definition equal). The system given in Eq.(\ref{cellSysMatApp2}) can thus not be solved independently for each grid-cell but must be solved for the entire grid.

In order to employ efficient techniques for solving linear systems, it is convenient to formulate the set of systems for each grid cell, Eq.(\ref{cellSysMatApp2}), into a single system for the entire grid, given in Eq.(\ref{gridLinSys}):
\begin{equation}
\label{gridLinSys}
\mathbf{\Pi} \mathbf{w} = \mathbf{t}
\end{equation}
where $\mathbf{\Pi}$ is a square, symmetric matrix and $\mathbf{w}$ and $\mathbf{t}$ are column vectors. This is done by defining a one to one index mapping from all the internal flow components to the components in the column vector, $\mathbf{w}$. Coefficients of the components in the grid cell systems (i.e. components in the symmetric matrix, $\mathbf{R}[k,l]$ in Eq.(\ref{cellSysMatApp2})), are added to $\mathbf{\Pi}$ if they correspond to internal components. If a component of $\mathbf{z}[k,l]$ is a boundary component, then it is multiplied with the corresponding row in $\mathbf{R}[k,l]$ and subtracted, forming the right hand vector, $\mathbf{t}$, in Eq.(\ref{gridLinSys}). This procedure is shown in detail by Algorithm \ref{assembleGrid}.

It is convenient to arrange components so that the matrix, $\mathbf{\Pi}$, gets a narrow band structure, allowing more efficient computations on the system. For computations presented in this paper the order is arranged by first sorting the internal flow components according to their location in the grid (indices $k,l$), then by what type of flow component ($x$-velocity, $y$-velocity or pressure), then by the order of the derivative (indices $\alpha,\beta$).

Note that Algorithm \ref{assembleGrid} shows the entire matrix, $\mathbf{\Pi}$, being assembled. In the implementation of this algorithm the sub-diagonal elements are not stored since the matrix is symmetric.

\begin{algorithm}
\begin{algorithmic}
\State $\mathbf{\Pi} \gets \mathbf{0}, \; \mathbf{t} \gets \mathbf{0}$
\For{$0 \leq k < L, \; 0 \leq l < L$}
	\For{$0 \leq m < 3N^2$}
           \State $r \gets \mathrm{index}(k,l,m)$
           \If{$0 \leq r$}
			\For{$0 \leq n < 3N^2$}
	            \State $c \gets \mathrm{index}(k,l,n)$
	            \If{$0 \leq c$}					
					\State $\mathbf{\Pi}[r,c] \gets \mathbf{\Pi}[r,c] + \mathbf{R}[k,l][m,n]$
				\Else
					\State $\mathbf{t}[r] \gets \mathbf{t}[r] - \mathbf{R}[k,l][m,n] \mathbf{z}[m,0]$
				\EndIf
			\EndFor
           \EndIf
	\EndFor
\EndFor
\end{algorithmic}
\caption{This pseudo-code shows the details of how the set of grid-cell systems (Eq.(\ref{cellSysMatApp2})) is reformulated into the system given by Eq.(\ref{gridLinSys}). The variables, $k$, $l$, $m$, $n$, $L$ and $N$ are all integers following their previous definitions from Section \ref{framework}. The temporary variables $r$ and $c$ are also integers and contain the row and column indices for the matrix $\mathbf{\Pi}$. Internal flow components are ordered, first by the grid point location, then by what type of flow component, and then by the order of the derivative, into a contiguous list. The method, index($\dots$), returns the position in this list if its arguments correspond to a an internal component. Otherwise, a negative number is returned.}
\label{assembleGrid}
\end{algorithm}

\section{Iterative Solution of the Nonlinear System of Equations}
\label{itsolve}
The Navier-Stokes equations on matrix form, Eq.(\ref{cellSys}), are solved by an iteration over several stages. Initially the internal flow components are either set to zero (velocity) and one (pressure) or corresponding to a solution for a lower Reynolds number, forming an initial approximate matrix, $\mathbf{\Pi}_0$, and an approximate solution, $\mathbf{w}_0$, of the system given in Eq.(\ref{gridLinSys}).
\subsection{Linear Substep}
\label{linearCG}
At the $(\kappa-1)$'th iteration, the system, Eq.(\ref{gridLinSys}), is formed using the approximate values, $\mathbf{w}_{\kappa-1}$. A new approximate solution, $\mathbf{w}'_{\kappa-1}$, is found using the linear conjugate gradient iteration \citep[see for example][chap. 38]{trefethen1997}. The linear conjugate gradient iteration is terminated when the relative improvement factor, $\hat{r}_{\kappa-1}$, defined in Eq.(\ref{linImp}), of the solution of the linear system reaches a predetermined value, $\hat{r}_{\kappa-1} \leq \hat{\omega} \ll 1$. 
\begin{equation}
\label{linImp}
\frac{\left\Vert \mathbf{\Pi}_{\kappa-1} \mathbf{w}'_{\kappa-1} - \mathbf{t}_{\kappa-1} \right\Vert_2 }{\left\Vert  \mathbf{\Pi}_{\kappa-1} \mathbf{w}_{\kappa-1} - \mathbf{t}_{\kappa-1} \right\Vert_2 } \equiv \hat{r}_{\kappa-1}
\end{equation}
The approximation, $\mathbf{w}'_{\kappa-1}$, is used to define a search direction, $\Delta \mathbf{w}_{\kappa-1}$, as shown in Eq.(\ref{searchGlob}):
\begin{equation}
\label{searchGlob}
\Delta \mathbf{w}_{\kappa-1} \equiv \mathbf{w}'_{\kappa-1} - \mathbf{w}_{\kappa-1}
\end{equation}
The search direction is defined to have corresponding grid-cell components given by Eq.(\ref{searchLoc}):
\begin{equation}
\label{searchLoc}
\Delta \mathbf{z}_{\kappa-1}[k,l] = \mathbf{z}'_{\kappa-1}[k,l] - \mathbf{z}_{\kappa-1}[k,l]
\end{equation}
where the mapping from $\mathbf{z}'_{\kappa-1}$ to $\mathbf{w}'_{\kappa-1}$ is the same as used in Algorithm \ref{assembleGrid} for internal components. If a component, $\mathbf{z}'_{\kappa-1}[k,l][\tau,0]$, corresponds to a boundary component, then it is defined to be equal to its initial value, $\mathbf{z}_{\kappa-1}[k,l][\tau,0]$, giving $\Delta \mathbf{z}_{\kappa-1}[k,l][\tau,0] = 0$ for boundary components. The updated flow components, $\mathbf{z}_{\kappa}[k,l]$, are given by Eq.(\ref{newLoc}):
\begin{equation}
\label{newLoc}
\left[ \begin{array}{c}
\mathbf{u}_{\kappa}[k,l] \\
\mathbf{v}_{\kappa}[k,l] \\
\mathbf{p}_{\kappa}[k,l] 
\end{array} \right]
= 
\left[ \begin{array}{c}
\mathbf{u}_{\kappa-1}[k,l] \\
\mathbf{v}_{\kappa-1}[k,l] \\
\mathbf{p}_{\kappa-1}[k,l] 
\end{array} \right] +
\left[ \begin{array}{c}
\theta_u \Delta \mathbf{u}_{\kappa-1}[k,l] \\
\theta_v \Delta \mathbf{v}_{\kappa-1}[k,l] \\
\theta_p \Delta \mathbf{p}_{\kappa-1}[k,l] 
\end{array} \right]
\end{equation}
where $\theta_u$, $\theta_v$ and $\theta_p$ are three parameters to be determined in each iterative step and 
\begin{equation}
\label{nonlinearDep}
\mathbf{z}_{\kappa}[k,l] \equiv 
\left[ \begin{array}{c}
\mathbf{u}_{\kappa}[k,l] \\
\mathbf{v}_{\kappa}[k,l] \\
\mathbf{p}_{\kappa}[k,l] 
\end{array} \right], \;
\mathbf{z}_{\kappa-1}[k,l] \equiv 
\left[ \begin{array}{c}
\mathbf{u}_{\kappa-1}[k,l] \\
\mathbf{v}_{\kappa-1}[k,l] \\
\mathbf{p}_{\kappa-1}[k,l] 
\end{array} \right], \;
\Delta \mathbf{z}_{\kappa-1}[k,l] \equiv 
\left[ \begin{array}{c}
\Delta \mathbf{u}_{\kappa-1}[k,l] \\
\Delta \mathbf{v}_{\kappa-1}[k,l] \\
\Delta \mathbf{p}_{\kappa-1}[k,l] 
\end{array} \right]
\end{equation}
in accordance with the definition given in Eq.(\ref{cellVec}). 
\subsection{Nonlinear Substep}
\label{nonlincg}
Consider the integrand of the grid-cell residual squared (Eq.(\ref{cellError})) at the $\kappa$'th stage:
\begin{equation}
\label{cellErrorIntegrand}
 \mathbf{z}_{\kappa}^T[k,l] \mathbf{E}_{\kappa}^T[k,l](x,y) \mathbf{E}_{\kappa}[k,l](x,y) \mathbf{z}_{\kappa}[k,l] 
\equiv \rho_\kappa[k,l](\theta_u,\theta_v,\theta_p,x,y)
\end{equation}
According to Eq.(\ref{newLoc}), the column vector, $\mathbf{z}_{\kappa}$, depends linearly on the parameters $\theta_u$, $\theta_v$ and $\theta_p$, and the matrix, $\mathbf{E}_{\kappa}(x,y)$, depends linearly on the parameters $\theta_u$ and $\theta_v$. Eq.(\ref{cellErrorIntegrand}) can thus be written as a fourth degree polynomial of $\theta_u$, $\theta_v$ and $\theta_p$, shown in Eq.(\ref{polyInteg})
\begin{align}
\label{polyInteg}
\rho_\kappa[k,l](\theta_u,\theta_v,\theta_p,x,y) = \, &c_{000} + c_{100} \theta_u + c_{200} \theta_u^2 + c_{300} \theta_u^3 + c_{400} \theta_u^4 + \nonumber \\  
&c_{010} \theta_v + c_{110} \theta_u  \theta_v + c_{210} \theta_u^2  \theta_v + c_{310} \theta_u^3 \theta_v + \nonumber \\
&c_{020} \theta_v^2 + c_{120} \theta_u  \theta_v^2 + c_{220} \theta_u^2   \theta_v^2 + \nonumber \\
&c_{030} \theta_v^3 + c_{130} \theta_u  \theta_v^3 + \\ 
&c_{040} \theta_v^4 + \nonumber \\
&c_{001} \theta_p + c_{101} \theta_u \theta_p + c_{201} \theta_u^2 \theta_p + \nonumber \\  
&c_{011} \theta_v \theta_p + c_{111} \theta_u \theta_v \theta_p + \nonumber \\
&c_{021} \theta_v^2 \theta_p + c_{002} \theta_p^2 \nonumber 
\end{align}
The coefficients, $c_{\dots}[k,l](x,y)$, in Eq.(\ref{polyInteg}) are determined from the definition of $\mathbf{E}[k,l](x,y)$ (see Eq.(\ref{cellMat})) and $\mathbf{m}[k,l](x,y)$ (see Eq.(\ref{diffDiscM})) through basic algebraic operations by substituting $\mathbf{u}_{\kappa-1} + \theta_u \Delta \mathbf{u}_{\kappa-1}$ for $\mathbf{u}$,  $\;\mathbf{v}_{\kappa-1} + \theta_v \Delta \mathbf{v}_{\kappa-1}$ for $\mathbf{v}$ and  $\mathbf{p}_{\kappa-1} + \theta_p \Delta \mathbf{p}_{\kappa-1}$ for $\mathbf{p}$ (the grid cell indices $[k,l]$ and function arguments $(x,y)$ for the coefficients, $c_{\dots}[k,l](x,y)$, are omitted in Eq.(\ref{polyInteg}) for the sake of brevity). Eq.(\ref{polyInteg}) is integrated numerically over the entire grid by the sum given in Eq.(\ref{coefInt}):
\begin{equation}
\label{coefInt}
P(\theta_u,\theta_v,\theta_p) \equiv \frac{1}{W} \sum_{\substack{k = 0 \\ l = 0}}^{L-2} \sum_{s = 0}^{S^2-1} \rho_\kappa[k,l](\theta_u,\theta_v,\theta_p,x_s,y_s)
\end{equation}
where $W = S^2 (L-1)^2$, yielding 
\begin{align}
\label{polyIntegrated}
P(\theta_u,\theta_v,\theta_p) = \, &C_{000} + C_{100} \theta_u + C_{200} \theta_u^2 + C_{300} \theta_u^3 + C_{400} \theta_u^4 + \nonumber \\  
&C_{010} \theta_v + C_{110} \theta_u  \theta_v + C_{210} \theta_u^2  \theta_v + C_{310} \theta_u^3 \theta_v + \nonumber \\
&C_{020} \theta_v^2 + C_{120} \theta_u  \theta_v^2 + C_{220} \theta_u^2   \theta_v^2 + \nonumber \\
&C_{030} \theta_v^3 + C_{130} \theta_u  \theta_v^3 + \\ 
&C_{040} \theta_v^4 + \nonumber \\
&C_{001} \theta_p + C_{101} \theta_u \theta_p + C_{201} \theta_u^2 \theta_p + \nonumber \\  
&C_{011} \theta_v \theta_p + C_{111} \theta_u \theta_v \theta_p + \nonumber \\
&C_{021} \theta_v^2 \theta_p + C_{002} \theta_p^2 \nonumber 
\end{align}
The function, $P(\theta_u,\theta_v,\theta_p)$, is then minimized with respect to the parameters $\theta_u,\theta_v,\theta_p$. The minimization of Eq.(\ref{polyIntegrated}) is not a computationally expensive step since the function, $P(\theta_u,\theta_v,\theta_p)$, is a fourth degree polynomial depending on only three variables. For the purposes of this paper, the nonlinear conjugate gradient iteration was sufficient. A fixed number of iterations (50) was used and the Fletcher-Reeves method determined the line search direction (the reader may refer to \citet{Shewchuk1994} for details concerning the nonlinear conjugate gradient iteration).

With the parameters $\theta_u,\theta_v,\theta_p$ determined, the flow components are updated as shown in Eq.(\ref{newLoc}) and the iteration may be repeated until desired accuracy is reached or until errors, due to limited floating point precision or due to the approximate nature of the discretization, prevents further improvement.

\section{Results}
Solutions were computed for Reynolds numbers, $\mathrm{Re} \in \{ 100, 1000, 5000, 10000, 20000, 30000, 40000 \}$. The data from all the computations is too extensive to be displayed in detail in this paper but is available from the author upon request. In Subsections \ref{velProf}-\ref{compTime} details of a selection of the computed solutions are discussed.


\label{results}
\subsection{Velocity Profiles}
\label{velProf}

\begin{figure}[h]
\centering
\setlength{\unitlength}{0.5\textwidth}
\begin{minipage}[c]{\unitlength}
\begin{picture}(1,1)
    \put(0,0){\includegraphics[width=\unitlength]{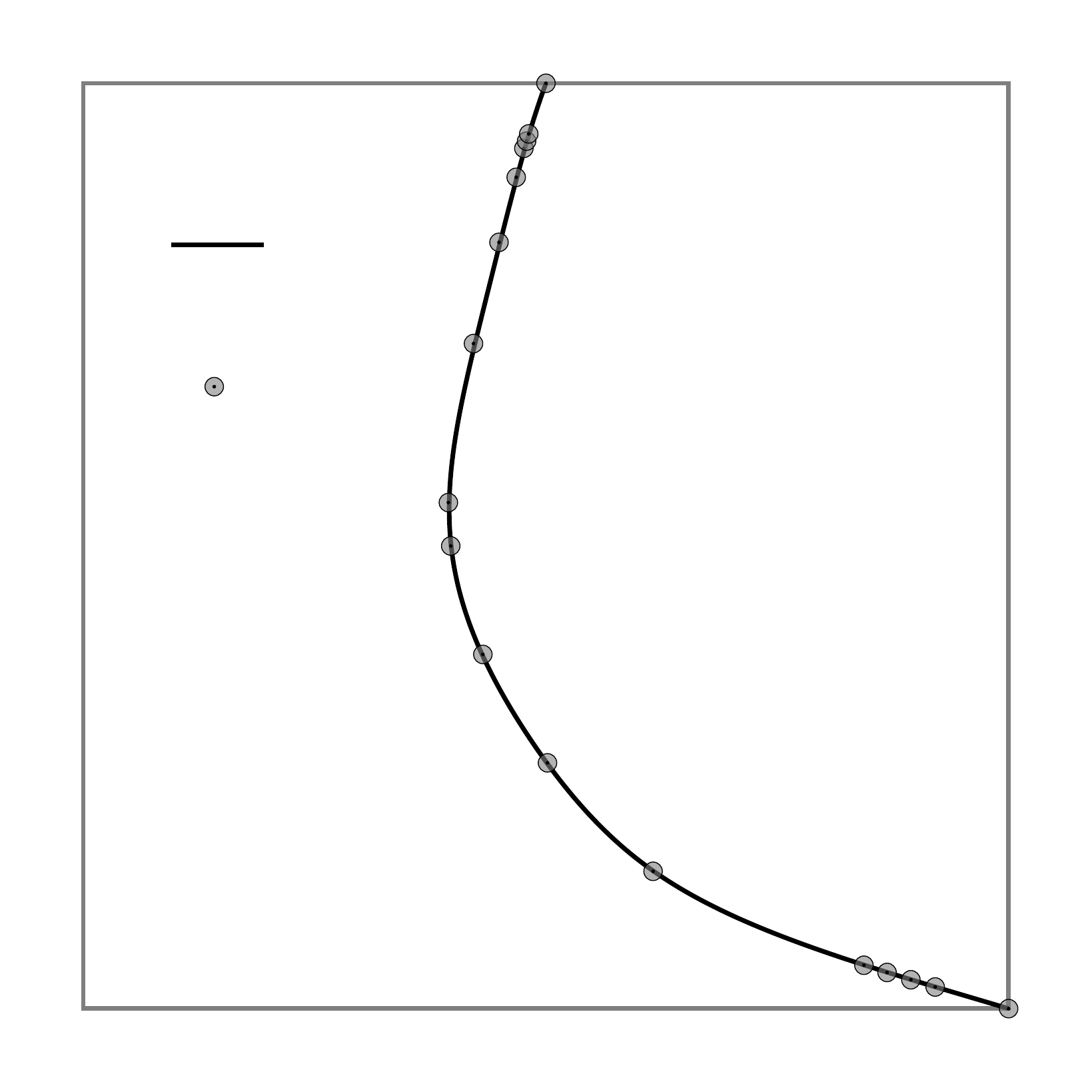}}
	\put(0.08,0.02){\color[rgb]{0,0,0} \makebox(0,0)[cb]{\smash{$-1$}}}
	\put(0.5,0.01){\color[rgb]{0,0,0} \makebox(0,0)[cb]{\smash{$u \; \longrightarrow$}}}
	\put(0.12,0.67){\color[rgb]{0,0,0} \makebox(0,0)[lb]{\smash{Ghia et al.}}}
	\put(0.12,0.80){\color[rgb]{0,0,0} \makebox(0,0)[lb]{\smash{Current}}}
	\put(0.92,0.02){\color[rgb]{0,0,0} \makebox(0,0)[cb]{\smash{$1$}}}
	\put(0.04,0.50){\color[rgb]{0,0,0}\rotatebox{90}{\makebox(0,0)[cb]{\smash{$y \rightarrow$}}}}
	\put(0.05,0.09){\color[rgb]{0,0,0}\rotatebox{90}{\makebox(0,0)[cb]{\smash{$0$}}}}
	\put(0.05,0.91){\color[rgb]{0,0,0}\rotatebox{90}{\makebox(0,0)[cb]{\smash{$4$}}}}
\end{picture}
\end{minipage} \; 
\begin{minipage}[c]{0.6\unitlength}
\caption{This figure shows the computed $x$-component of the velocity on a vertical line through the geometric center of the grid for Reynolds number, $\mathrm{Re} = 100$. The line shows current results, $\mathbf{b}(x',y') \mathbf{B}^{-1} \mathbf{u}[k,l]$ for $x = (L-1)/2$, $0 \leq y \leq L-1$, with $L = 5$ and $\Omega = 4$ where $x' = x - \mathbf{x}\left[k,l\right]$ and $y' = y - \mathbf{y}\left[k,l\right]$. The dotted circles show results presented by \citet{Ghia82} as a comparison.}
\label{re100xslice}
\end{minipage}
\end{figure}
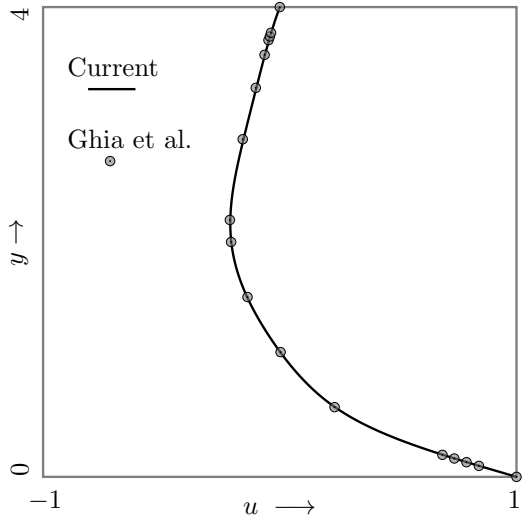

\begin{figure}[h]
\centering
\setlength{\unitlength}{0.48\textwidth}
\subfigure[The complete velocity profile]{
   \begin{picture}(1,1)
   \put(0,0){\includegraphics[width=\unitlength]{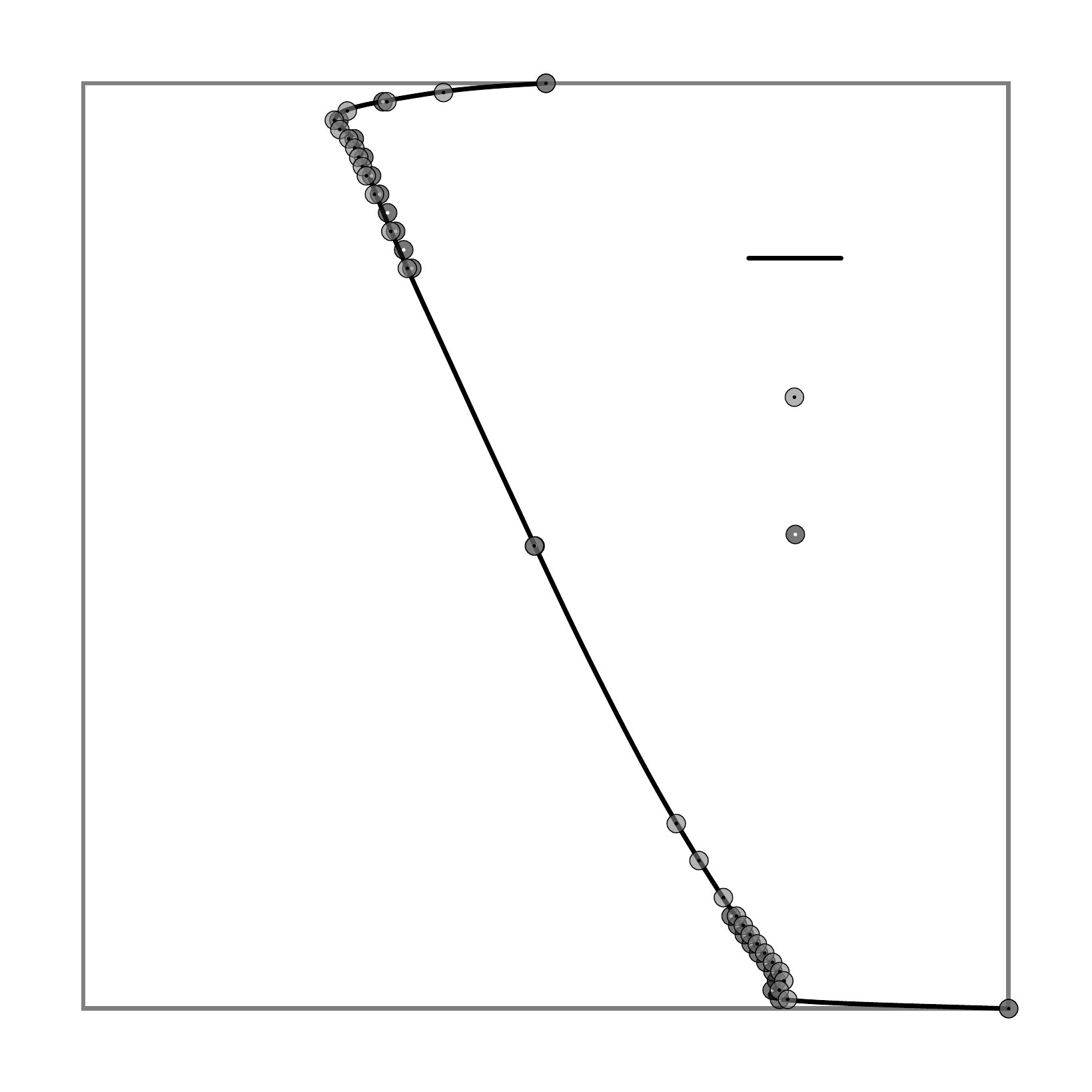}}
   \put(0.08,0.02){\color[rgb]{0,0,0} \makebox(0,0)[cb]{\smash{$-1$}}}
	\put(0.5,0.01){\color[rgb]{0,0,0} \makebox(0,0)[cb]{\smash{$u \; \longrightarrow$}}}
	\put(0.74,0.66){\color[rgb]{0,0,0} \makebox(0,0)[cb]{\smash{Wahba}}}
	\put(0.74,0.54){\color[rgb]{0,0,0} \makebox(0,0)[cb]{\smash{Erturk et al.}}}
	\put(0.74,0.79){\color[rgb]{0,0,0} \makebox(0,0)[cb]{\smash{Current}}}
	\put(0.92,0.02){\color[rgb]{0,0,0} \makebox(0,0)[cb]{\smash{$1$}}}
	\put(0.04,0.50){\color[rgb]{0,0,0}\rotatebox{90}{\makebox(0,0)[cb]{\smash{$y \rightarrow$}}}}
	\put(0.05,0.09){\color[rgb]{0,0,0}\rotatebox{90}{\makebox(0,0)[cb]{\smash{$0$}}}}
	\put(0.05,0.91){\color[rgb]{0,0,0}\rotatebox{90}{\makebox(0,0)[cb]{\smash{$99$}}}}
   \end{picture}
   \label{re20000sliceFull}
   }
\subfigure[The upper and lower range of the velocity profile]{
   \begin{picture}(1,1)
   \put(0,0){\includegraphics[width=\unitlength]{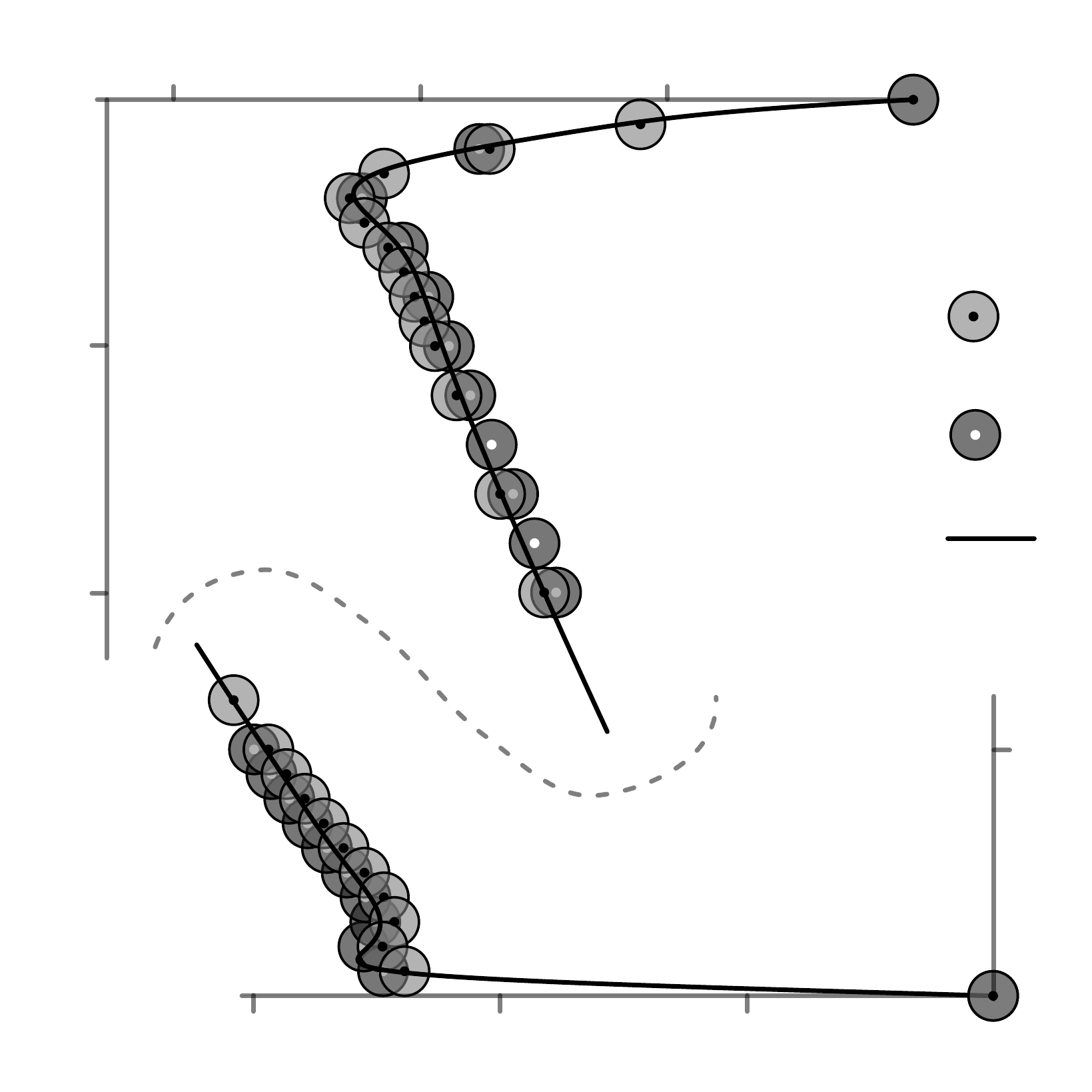}}
	\put(0.99,0.09){\color[rgb]{0,0,0}\rotatebox{90}{\makebox(0,0)[cb]{\smash{$0$}}}}
	\put(0.99,0.31){\color[rgb]{0,0,0}\rotatebox{90}{\makebox(0,0)[cb]{\smash{$\frac{1}{10}99$}}}}
	\put(0.05,0.91){\color[rgb]{0,0,0}\rotatebox{90}{\makebox(0,0)[cb]{\smash{$99$}}}}
	\put(0.05,0.67){\color[rgb]{0,0,0}\rotatebox{90}{\makebox(0,0)[cb]{\smash{$\frac{9}{10}99$}}}}
	\put(0.05,0.45){\color[rgb]{0,0,0}\rotatebox{90}{\makebox(0,0)[cb]{\smash{$\frac{8}{10}99$}}}}
	\put(0.912,0.02){\color[rgb]{0,0,0} \makebox(0,0)[cb]{\smash{$1$}}}
	\put(0.69,0.02){\color[rgb]{0,0,0} \makebox(0,0)[cb]{\smash{$0.8$}}}
	\put(0.46,0.02){\color[rgb]{0,0,0} \makebox(0,0)[cb]{\smash{$0.6$}}}
	\put(0.231,0.02){\color[rgb]{0,0,0} \makebox(0,0)[cb]{\smash{$0.4$}}}

	\put(0.837,0.95){\color[rgb]{0,0,0} \makebox(0,0)[cb]{\smash{$0$}}}
	\put(0.60,0.95){\color[rgb]{0,0,0} \makebox(0,0)[cb]{\smash{$-0.2$}}}
	\put(0.37,0.95){\color[rgb]{0,0,0} \makebox(0,0)[cb]{\smash{$-0.4$}}}
	\put(0.142,0.95){\color[rgb]{0,0,0} \makebox(0,0)[cb]{\smash{$-0.6$}}}

	\put(0.86,0.70){\color[rgb]{0,0,0} \makebox(0,0)[rb]{\smash{Wahba}}}
	\put(0.86,0.59){\color[rgb]{0,0,0} \makebox(0,0)[rb]{\smash{Erturk et al.}}}
	\put(0.86,0.49){\color[rgb]{0,0,0} \makebox(0,0)[rb]{\smash{Current}}}

    \put(0.27,-0.01){\color[rgb]{0,0,0} \makebox(0,0)[lb]{\smash{$u \; \longrightarrow$}}}

	\put(0.04,0.2){\color[rgb]{0,0,0}\rotatebox{90}{\makebox(0,0)[lb]{\smash{$y \rightarrow$}}}}
   \end{picture}
   \label{re20000sliceClose}
}
\caption{These figures show the computed $x$-component of the velocity on a vertical line through the geometric center of the grid for Reynolds number, $\mathrm{Re} = 20000$. The line shows current results, $\mathbf{b}(x',y') \mathbf{B}^{-1} \mathbf{u}[k,l]$ for $x = (L-1)/2$, $0 \leq y \leq L-1$, with $L = 100$ and $\Omega = 5$ where $x' = x - \mathbf{x}\left[k,l\right]$ and $y' = y - \mathbf{y}\left[k,l\right]$. The dotted circles show results presented by \citet{Erturk05} and \citet{Wahba12} as a comparison. Subfigure \ref{re20000sliceFull} shows the plot for the entire $y$-range while subfigure \ref{re20000sliceClose} shows a larger view of the upper and lower $y$-range.}
\label{re20000slice}
\end{figure}

\begin{figure}[h]
\centering
\setlength{\unitlength}{0.5\textwidth}
\begin{minipage}[c]{\unitlength}
\begin{picture}(1,1)
    \put(0,0){\includegraphics[width=\unitlength]{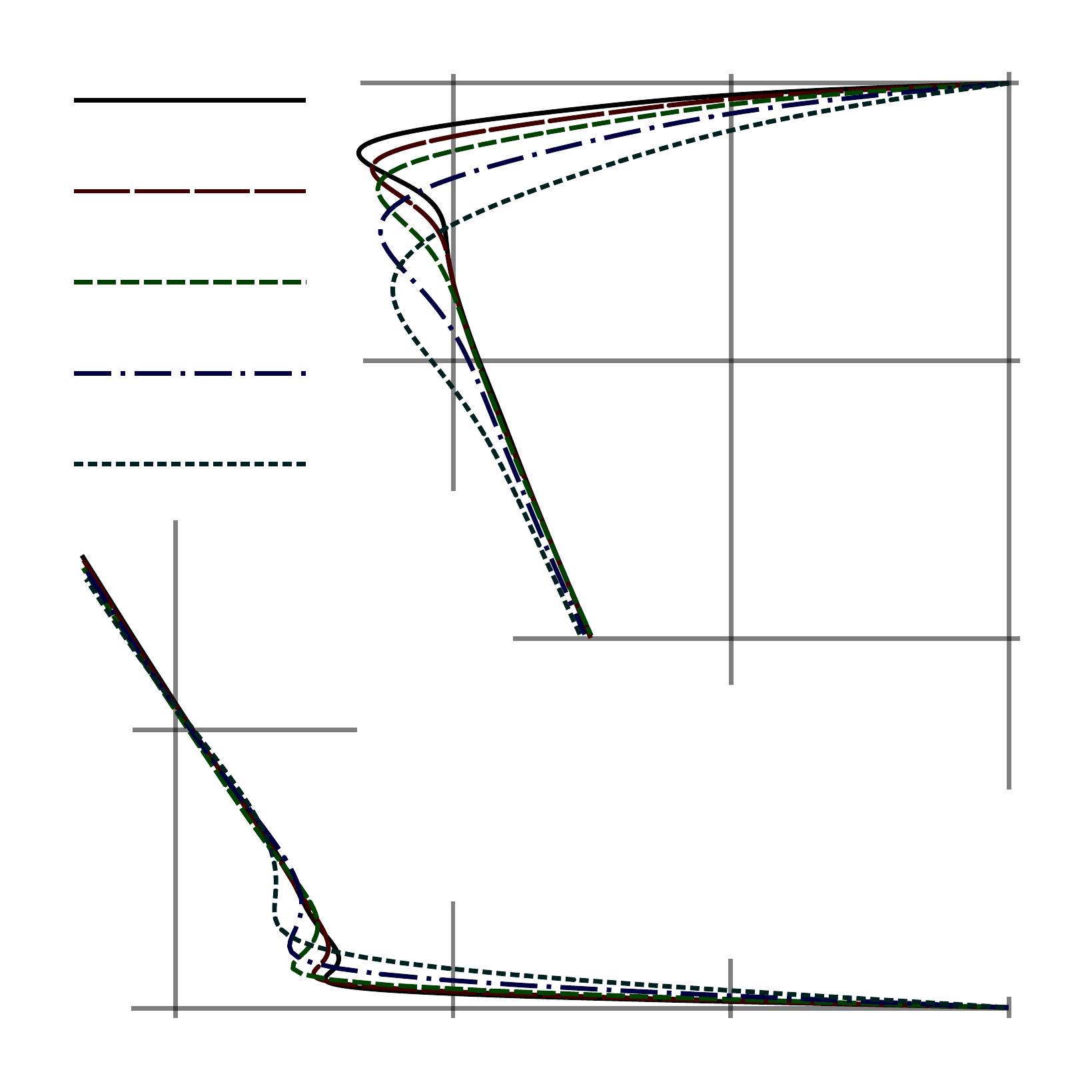}}
	\put(0.47,0.00){\color[rgb]{0,0,0} \makebox(0,0)[lb]{\smash{$u \; \longrightarrow$}}}
	\put(0.28,0.925){\color[rgb]{0,0,0} \makebox(0,0)[rb]{\smash{$\mathrm{Re} = 40000$}}}
	\put(0.28,0.840){\color[rgb]{0,0,0} \makebox(0,0)[rb]{\smash{$\mathrm{Re} = 30000$}}}
	\put(0.28,0.755){\color[rgb]{0,0,0} \makebox(0,0)[rb]{\smash{$\mathrm{Re} = 20000$}}}
	\put(0.28,0.670){\color[rgb]{0,0,0} \makebox(0,0)[rb]{\smash{$\mathrm{Re} = 10000$}}}
	\put(0.28,0.585){\color[rgb]{0,0,0} \makebox(0,0)[rb]{\smash{$\mathrm{Re} = 5000$}}}

	\put(0.04,0.50){\color[rgb]{0,0,0}\rotatebox{90}{\makebox(0,0)[cb]{\smash{$y \rightarrow$}}}}
	\put(0.07,0.06){\color[rgb]{0,0,0}\rotatebox{0}{\makebox(0,0)[rb]{\smash{$0$}}}}
	\put(0.07,0.32){\color[rgb]{0,0,0}\rotatebox{0}{\makebox(0,0)[rb]{\smash{$\frac{L-1}{10}$}}}}

	\put(0.925,0.02){\color[rgb]{0,0,0} \makebox(0,0)[cb]{\smash{$1$}}}
	\put(0.670,0.02){\color[rgb]{0,0,0} \makebox(0,0)[cb]{\smash{$0.8$}}}
	\put(0.415,0.02){\color[rgb]{0,0,0} \makebox(0,0)[cb]{\smash{$0.6$}}}
	\put(0.160,0.02){\color[rgb]{0,0,0} \makebox(0,0)[cb]{\smash{$0.4$}}}

	\put(0.925,0.95){\color[rgb]{0,0,0} \makebox(0,0)[cb]{\smash{$0$}}}
	\put(0.655,0.95){\color[rgb]{0,0,0} \makebox(0,0)[cb]{\smash{$-0.2$}}}
	\put(0.400,0.95){\color[rgb]{0,0,0} \makebox(0,0)[cb]{\smash{$-0.4$}}}

	\put(0.95,0.910){\color[rgb]{0,0,0}\rotatebox{0}{\makebox(0,0)[lb]{\smash{$L-1$}}}}
	\put(0.95,0.660){\color[rgb]{0,0,0}\rotatebox{0}{\makebox(0,0)[lb]{\smash{$\frac{9(L-1)}{10}$}}}}
	\put(0.95,0.403){\color[rgb]{0,0,0}\rotatebox{0}{\makebox(0,0)[lb]{\smash{$\frac{8(L-1)}{10}$}}}}
\end{picture}
\end{minipage} \hspace{0.1\unitlength} 
\begin{minipage}[c]{0.7\unitlength}
\caption{This figure shows the computed $x$-component of the velocity on a vertical line through the geometric center of the grid for Reynolds number, $\mathrm{Re} \in \{5000,10000,20000,30000,40000\}$. Only the upper ($y$ near $L-1$) and lower ($y$ near $0$) part of the computational domain is plotted. The different lines shows current results, $\mathbf{b}(x',y') \mathbf{B}^{-1} \mathbf{u}[k,l]$ for $x = (L-1)/2$, $0 \leq y \leq L-1$, with $L \in \{40,60,100,120,135\}$ and $\Omega = 5$ where $x' = x - \mathbf{x}\left[k,l\right]$ and $y' = y - \mathbf{y}\left[k,l\right]$.}
\label{highReComp}
\end{minipage}
\end{figure}

The $x$-component of the velocity, $u(x,y)$, through the geometric center of the cavity, from the center of the "lid", ($x = (L-1)/2, y = 0$), to the opposing side, ($x = (L-1)/2, y = L-1$), is shown in Figures \ref{re100xslice}-\ref{highReComp}. This will simply be referred to as a \textit{velocity profile} from now on. 

For Reynolds number, $\mathrm{Re} = 100$, the well known results from \citet{Ghia82} are used as a comparison (Figure \ref{re100xslice}). For Reynolds number, $\mathrm{Re} = 20000$, the current results are compared\footnote{Reynolds number, $\mathrm{Re} = 20000$, was the highest Reynolds number for which multiple reference results were available.} with the very fine-grid solutions from \citet{Erturk05} and \citet{Wahba12} (Figure \ref{re20000slice}). Additionally, the interesting features for various high Reynolds numbers from $\mathrm{Re} = 5000$ to $\mathrm{Re} = 40000$ is compared with each other (Figure \ref{highReComp}).

Figure \ref{re100xslice}, which shows the velocity profile for Reynolds number, $\mathrm{Re} = 100$, confirms that the current results agree with established results \citep{Ghia82} for this Reynolds number.

Figure \ref{re20000slice}, which shows the velocity profile for Reynolds number, $\mathrm{Re} = 20000$, shows a small deviance from the reference solutions by \citet{Erturk05} and \citet{Wahba12}. In this case the current solution tends to agree with the references where they coincide and tends to lie between the references where they do not coincide.

Figure \ref{highReComp} shows how the upper and lower parts of the velocity profile evolve as the Reynolds number increases. The upper part shows a systematic trend where minimum drops while shifting increasingly closer to the edge. The lower part shows a similar trend for $\mathrm{Re} = 5000$ to $\mathrm{Re} = 20000$. However from $\mathrm{Re} = 20000$ to $\mathrm{Re} = 40000$, the local minimum move toward the edge at a much smaller rate while increasing in magnitude at a greater rate.

Figure \ref{velPlots} shows the velocity profiles for Reynolds number, $\mathrm{Re} = 1000$, obtained with increasing values of $\Omega$ and decreasing values of $L$, compared with the results given by \citet{Ghia82,Erturk05}. The value of $L$ was the lowest value which did not give any significant deviance from solutions obtained using a higher resolution. These results illustrate how an increase of the order of continuity allows for a lower grid resolution while still achieving results of similar accuracy. Also note that the amount of data contained in the grid (proportional to $L^2 \Omega^2$ in a two dimensional grid) decreases with increasing values of $\Omega$. 

\begin{figure}[h]
\centering
\setlength{\unitlength}{0.98\textwidth}
\begin{minipage}[c]{\unitlength}
\begin{picture}(1,0.56)
    \put(0,0){\includegraphics[width=\unitlength]{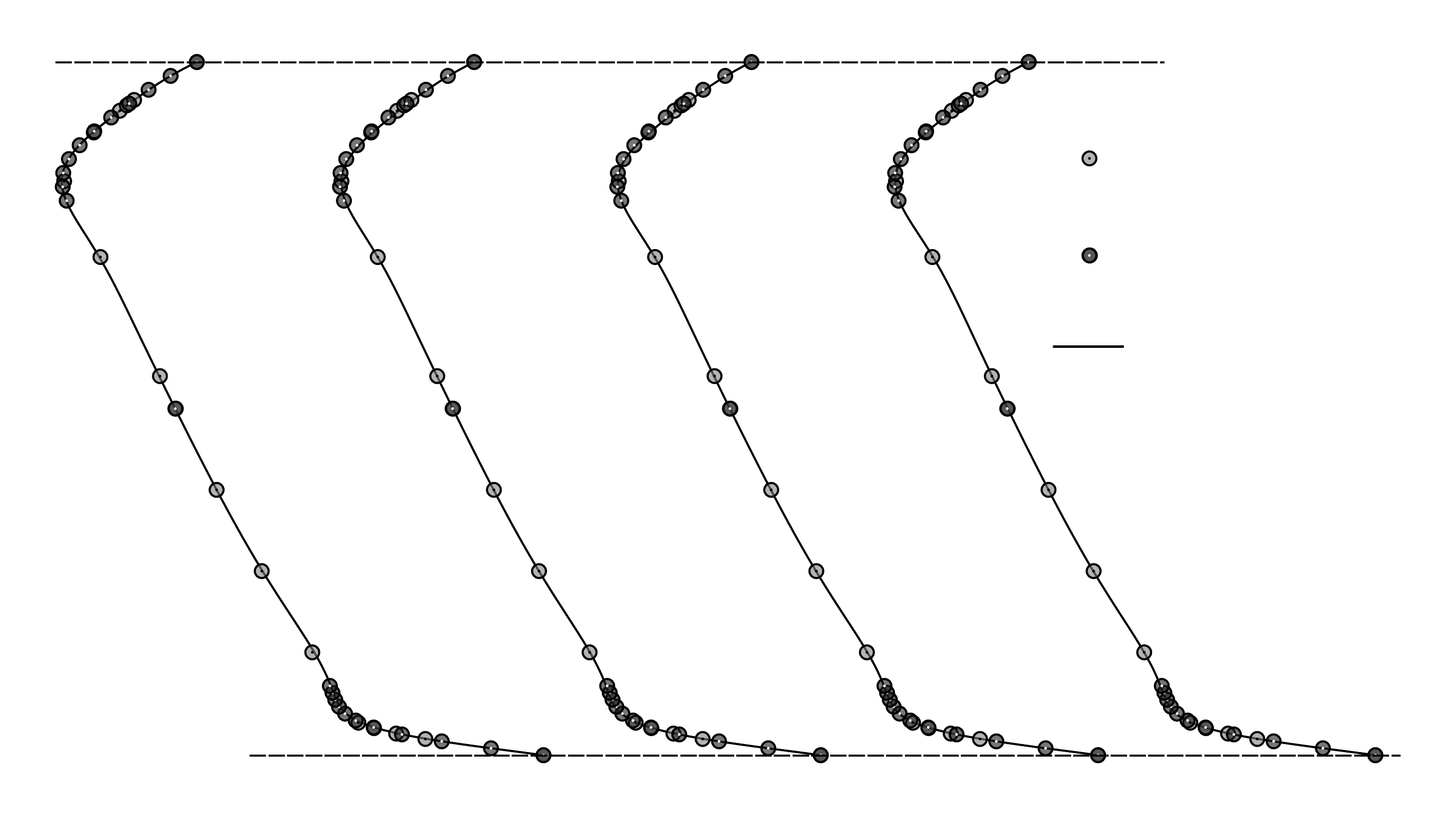}}
       \put(0.12,0.57){\color[rgb]{0,0,0} \makebox(0,0)[cb]{\smash{$L = 70$, $\Omega = 2$}}}
       \put(0.12,0.54){\color[rgb]{0,0,0} \makebox(0,0)[cb]{\smash{$L^2 \Omega^2 = 19600$}}}
       
       \put(0.32,0.57){\color[rgb]{0,0,0} \makebox(0,0)[cb]{\smash{$L = 35$, $\Omega = 3$}}}
       \put(0.32,0.54){\color[rgb]{0,0,0} \makebox(0,0)[cb]{\smash{$L^2 \Omega^2 = 11025$}}}
              
       \put(0.52,0.57){\color[rgb]{0,0,0} \makebox(0,0)[cb]{\smash{$L = 11$, $\Omega = 4$}}}
       \put(0.52,0.54){\color[rgb]{0,0,0} \makebox(0,0)[cb]{\smash{$L^2 \Omega^2 = 1936$}}}
       
       \put(0.72,0.57){\color[rgb]{0,0,0} \makebox(0,0)[cb]{\smash{$L = 8$, $\Omega = 5$}}}
       \put(0.72,0.54){\color[rgb]{0,0,0} \makebox(0,0)[cb]{\smash{$L^2 \Omega^2 = 1600$}}}
       
    \put(0.87,0.52){\color[rgb]{0,0,0} \makebox(0,0)[cb]{\smash{$y = L-1$}}}
   \put(0.13,0.036){\color[rgb]{0,0,0} \makebox(0,0)[cb]{\smash{$y = 0$}}}
    \put(0.76,0.47){\color[rgb]{0,0,0} \makebox(0,0)[cb]{\smash{Ghia et al.}}}
    \put(0.76,0.41){\color[rgb]{0,0,0} \makebox(0,0)[cb]{\smash{Erturk et al.}}}
    \put(0.76,0.34){\color[rgb]{0,0,0} \makebox(0,0)[cb]{\smash{Current}}}
\end{picture}
\end{minipage} \hspace{0.15\unitlength} 
\begin{minipage}[c]{\unitlength}
\caption{This figure shows a comparison of the computed $x$-component of the velocity on a vertical line through the geometric center of the grid for $(L,\Omega) \in \{(70,2),(35,3),(11,4),(8,5)\}$ with Reynolds number, $\mathrm{Re} = 1000$. At the top ($y = L-1$) of the figure the $x$-component of the velocity is zero and at the bottom ($y = 0$) it is one (positive $x$-direction). The dotted circles show results presented by \citet{Erturk05} and \citet{Ghia82}.}
\label{L_Omega_comp}
\end{minipage}
\end{figure}

\subsection{Flow Configurations}

\begin{figure}[h]
\centering
\setlength{\unitlength}{0.46\textwidth}
\subfigure[$L = 20$, $\Omega = 2$, $\mathrm{Re} = 100$]{
   \begin{picture}(1,1)
   \put(0,0){\includegraphics[width=\unitlength]{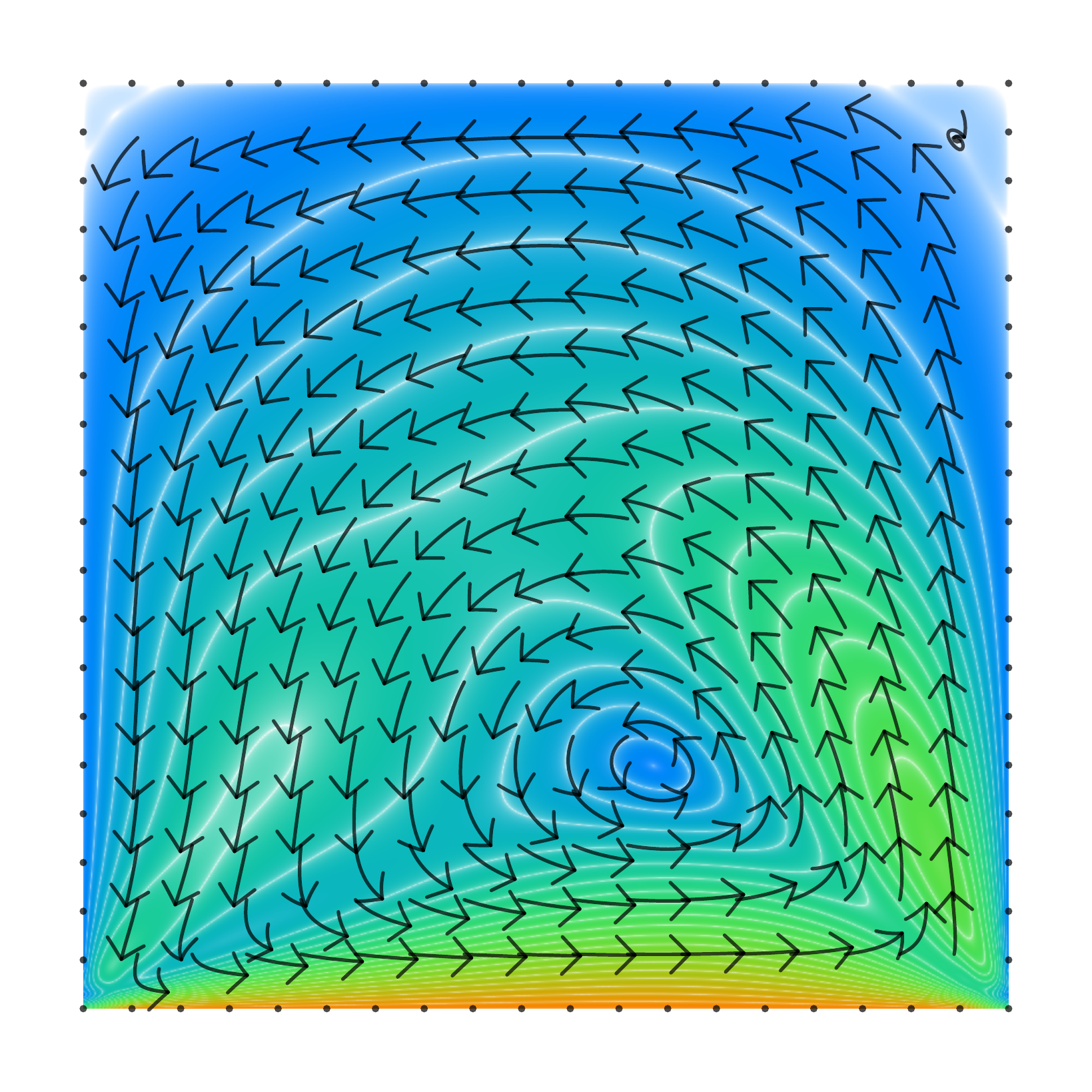}}
   \put(0.04,0.50){\color[rgb]{0,0,0}\rotatebox{90}{\makebox(0,0)[cb]{\smash{$y \rightarrow$}}}}
   \put(0.50,0.03){\color[rgb]{0,0,0}\makebox(0,0)[cb]{\smash{$x \rightarrow$}}}
   \put(0.08,0.02){\color[rgb]{0,0,0}\rotatebox{0}{\makebox(0,0)[cb]{\smash{$0$}}}}
   \put(0.92,0.02){\color[rgb]{0,0,0}\rotatebox{0}{\makebox(0,0)[cb]{\smash{$19$}}}}
   \put(0.05,0.08){\color[rgb]{0,0,0}\rotatebox{90}{\makebox(0,0)[cb]{\smash{$0$}}}}
   \put(0.05,0.92){\color[rgb]{0,0,0}\rotatebox{90}{\makebox(0,0)[cb]{\smash{$19$}}}}
   \end{picture}
   \label{velRe100full}
}
\subfigure[$L = 20$, $\Omega = 2$, $\mathrm{Re} = 100$]{
   \begin{picture}(1,1)
   \put(0,0){\includegraphics[width=\unitlength]{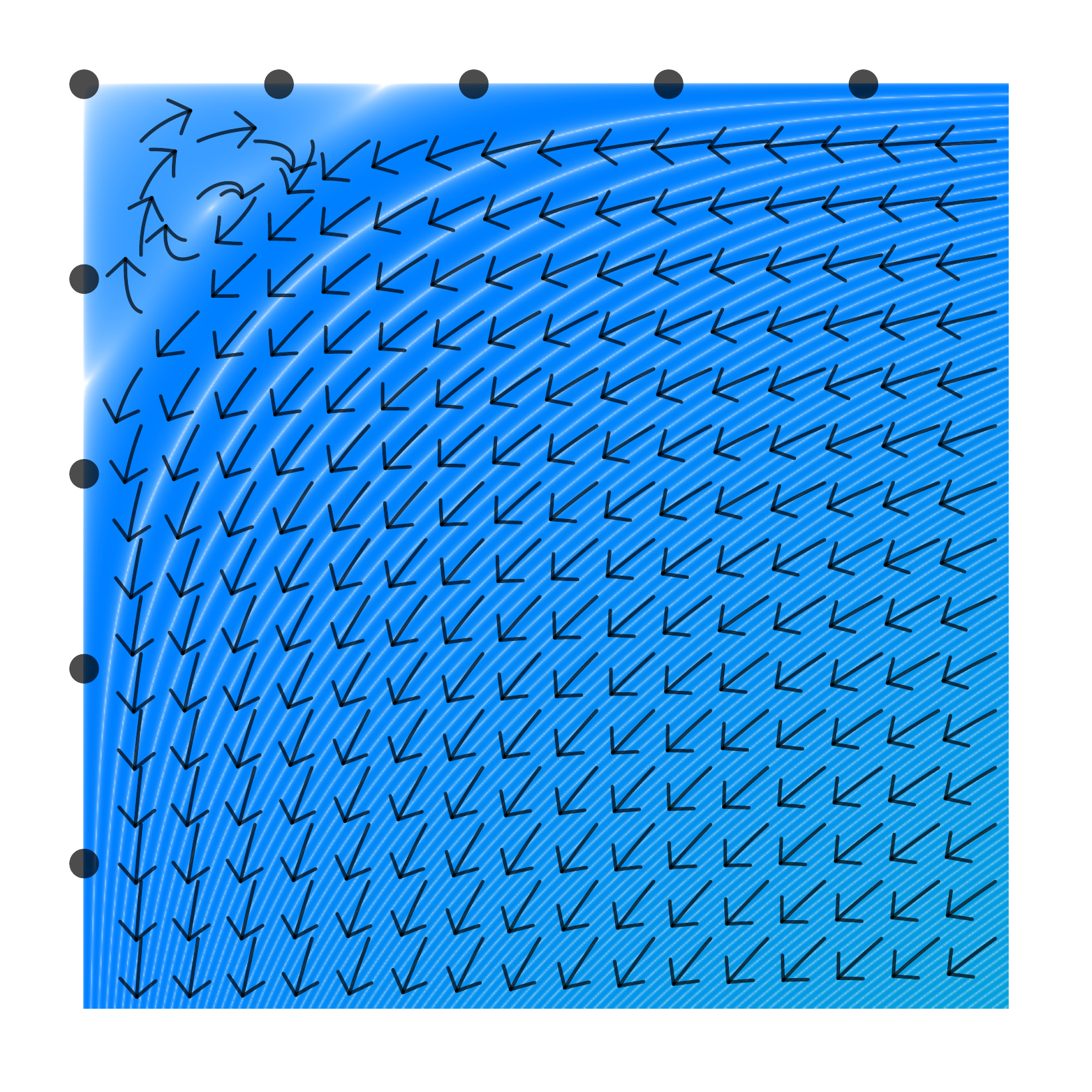}}
   \put(0.97,0.50){\color[rgb]{0,0,0}\rotatebox{90}{\makebox(0,0)[cb]{\smash{$y \rightarrow$}}}}
   \put(0.50,0.03){\color[rgb]{0,0,0}\makebox(0,0)[cb]{\smash{$x \rightarrow$}}}
   \put(0.05,0.92){\color[rgb]{0,0,0}\rotatebox{90}{\makebox(0,0)[cb]{\smash{$19$}}}}
   \put(0.05,0.565){\color[rgb]{0,0,0}\rotatebox{90}{\makebox(0,0)[cb]{\smash{$17$}}}}
   \put(0.05,0.205){\color[rgb]{0,0,0}\rotatebox{90}{\makebox(0,0)[cb]{\smash{$15$}}}}
   \put(0.080,0.95){\color[rgb]{0,0,0}\rotatebox{0}{\makebox(0,0)[cb]{\smash{$0$}}}}
   \put(0.435,0.95){\color[rgb]{0,0,0}\rotatebox{0}{\makebox(0,0)[cb]{\smash{$2$}}}}
   \put(0.790,0.95){\color[rgb]{0,0,0}\rotatebox{0}{\makebox(0,0)[cb]{\smash{$4$}}}}
   \end{picture}
   \label{velRe100close}
}
\subfigure[$L = 135$, $\Omega = 5$, $\mathrm{Re} = 40000$]{
   \begin{picture}(1,1)
   \put(0,0){\includegraphics[width=\unitlength]{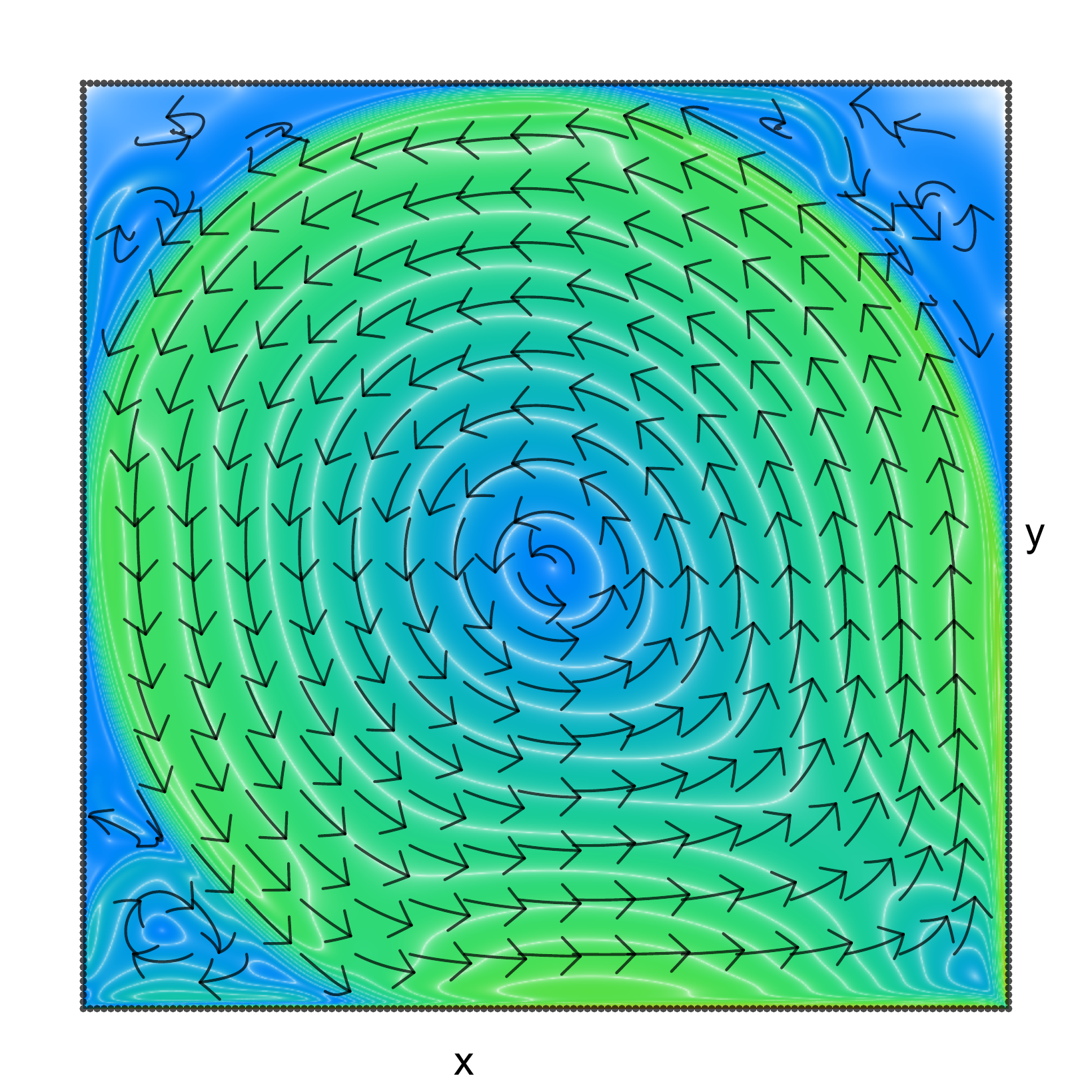}}
   \put(0.04,0.50){\color[rgb]{0,0,0}\rotatebox{90}{\makebox(0,0)[cb]{\smash{$y \rightarrow$}}}}
   \put(0.50,0.03){\color[rgb]{0,0,0}\makebox(0,0)[cb]{\smash{$x \rightarrow$}}}
   \put(0.08,0.02){\color[rgb]{0,0,0}\rotatebox{0}{\makebox(0,0)[cb]{\smash{$0$}}}}
   \put(0.92,0.02){\color[rgb]{0,0,0}\rotatebox{0}{\makebox(0,0)[cb]{\smash{$134$}}}}
   \put(0.05,0.08){\color[rgb]{0,0,0}\rotatebox{90}{\makebox(0,0)[cb]{\smash{$0$}}}}
   \put(0.05,0.92){\color[rgb]{0,0,0}\rotatebox{90}{\makebox(0,0)[cb]{\smash{$134$}}}}
   \end{picture}
   \label{velRe40000full}
}
\subfigure[$L = 135$, $\Omega = 5$, $\mathrm{Re} = 40000$]{
   \begin{picture}(1,1)
   \put(0,0){\includegraphics[width=\unitlength]{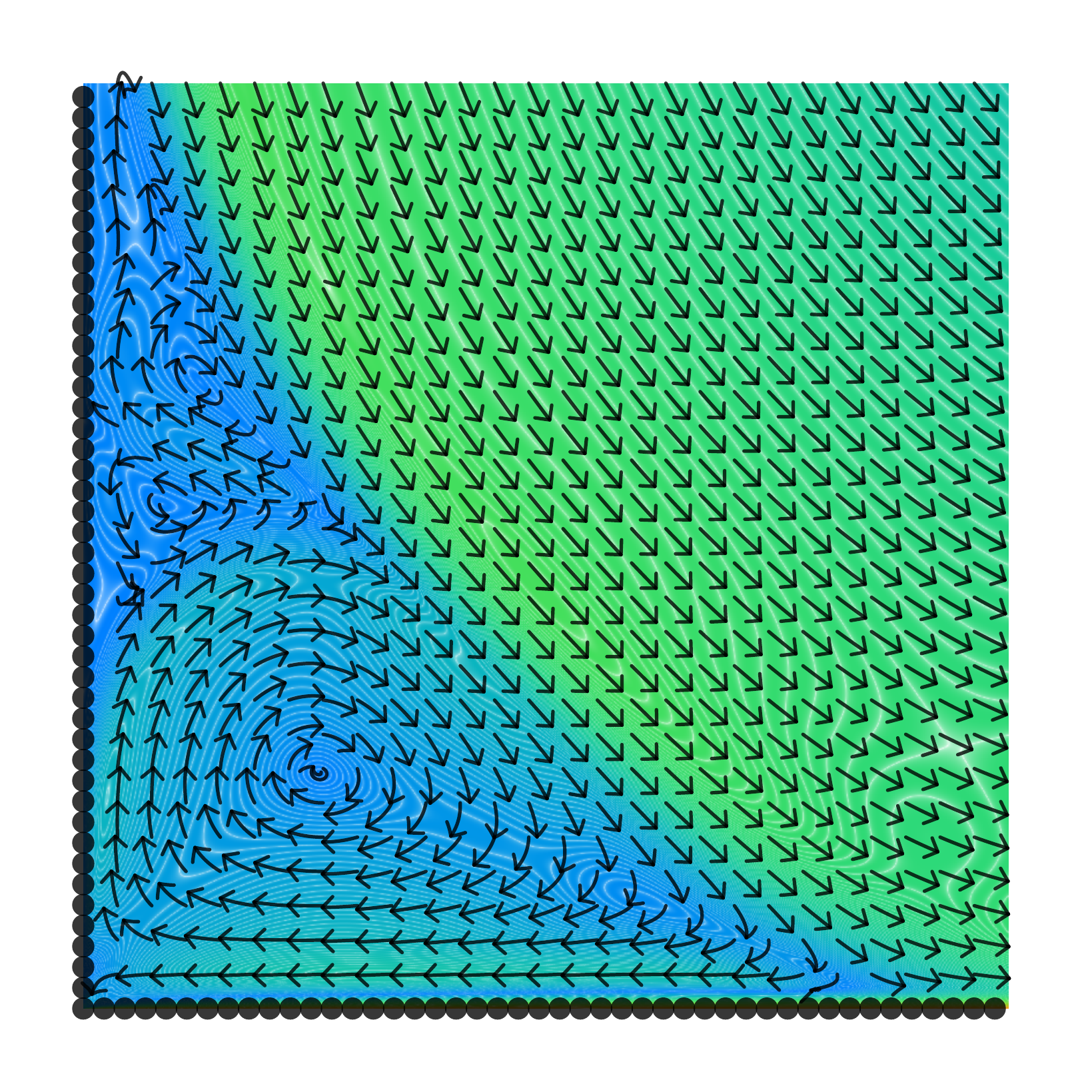}}
   \put(0.04,0.50){\color[rgb]{0,0,0}\rotatebox{90}{\makebox(0,0)[cb]{\smash{$y \rightarrow$}}}}
   \put(0.50,0.02){\color[rgb]{0,0,0}\makebox(0,0)[cb]{\smash{$x \rightarrow$}}}
   \put(0.05,0.908){\color[rgb]{0,0,0}\rotatebox{90}{\makebox(0,0)[cb]{\smash{$45$}}}}
   \put(0.05,0.074){\color[rgb]{0,0,0}\rotatebox{90}{\makebox(0,0)[cb]{\smash{$0$}}}}
   \put(0.075,0.02){\color[rgb]{0,0,0}\rotatebox{0}{\makebox(0,0)[cb]{\smash{$0$}}}}
   \put(0.910,0.02){\color[rgb]{0,0,0}\rotatebox{0}{\makebox(0,0)[cb]{\smash{$45$}}}}
   \end{picture}
   \label{velRe40000close}
}
\caption{Subfigures \ref{velRe100full} and \ref{velRe100close} show the computed flow configuration for Reynolds number, $\mathrm{Re} = 100$, with $L = 20$ and $\Omega = 2$. Subfigure \ref{velRe100full} shows the entire computational domain, while Subfigure \ref{velRe100close} shows details at the upper left corner (commonly referred to as the tertiary vortex). Subfigures \ref{velRe40000full} and \ref{velRe40000close} show the computed flow configuration for Reynolds number, $\mathrm{Re} = 40000$, with $L = 135$ and $\Omega = 5$. Subfigure \ref{velRe40000full} shows the entire computational domain, while Subfigure \ref{velRe40000close} shows details at the lower left corner (commonly referred to as the quaternary vortex, however in this case it is split up into multiple sub-vortices). The color indicates the magnitude of the velocity, $\Vert (u,v) \Vert_2$, where orange is for $\Vert (u,v) \Vert_2 = 1$, green is for $\Vert (u,v) \Vert_2 = 1/2$ and bright blue is for $\Vert (u,v) \Vert_2 = 0$ (and interpolated between these colors for the intermediate values). Contour lines of the velocity magnitude are drawn in white with a contour interval of $1/20$ in Subfigure \ref{velRe100full},  $1/1000$ in Subfigure \ref{velRe100close}, $1/20$ in Subfigure \ref{velRe40000full} and $1/200$ in Subfigure \ref{velRe40000close}. The arrows are of constant length in each subfigure and are drawn in a Lagrangian coordinate system, defined by the two orthogonal unit vectors $\hat{x}_l = (u,v)/\Vert (u,v) \Vert_2$ and $\hat{y}_l = (v,-u)/\Vert (u,v) \Vert_2$, pointing in the positive direction along the unit vector, $\hat{x}_l$. The grid resolution is indicated by dots along the edge of the grid.}
\label{velPlots}
\end{figure}

Figure \ref{velPlots} shows visualizations of computed flow configurations. Due to the high polynomial degree of the solution, flow features below grid resolution are resolved. This can be seen in the close-up plot in Subfigure \ref{velRe100close}. At higher Reynolds numbers the required grid resolution is higher compared to the scale of the main vortices of the flow. However, secondary, tertiary and quaternary vortices all split up in several sub-vortices at high Reynolds numbers. It seems reasonable to assume that the higher resolution requirement is connected to this phenomenon.

\subsection{Computation Time and Convergence}
\label{compTime}
The computation times were achieved on a standard desktop computer (quad core Xeon W3565 CPU at 3.2 GHz with 12 GB RAM). The computation times are not comparable to what might be achieved on a high end system utilizing parallel computing, but might have some use for internal comparison. Up to four separate computations were run simultaneously (each single threaded) each utilizing 23-25 percent of the CPU capacity. 

\begin{figure}[h]
\centering
\setlength{\unitlength}{0.45\textwidth}

\subfigure[$L = 20$, $\Omega = 3$, $\mathrm{Re} = 1000$]{
   \begin{picture}(1,1)
   \put(0,0){\includegraphics[width=\unitlength]{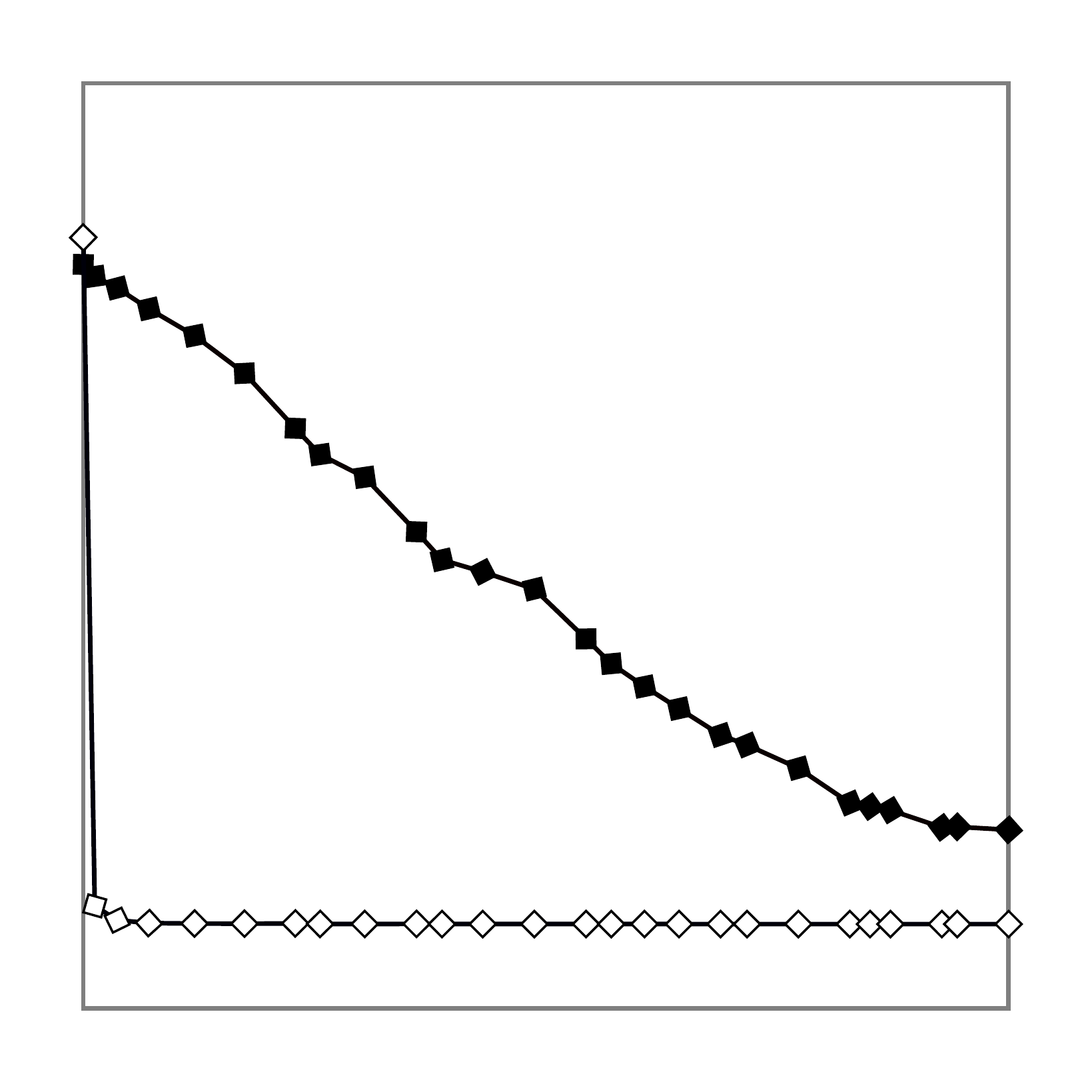}}
   \put(0.580,0.510){\color[rgb]{0,0,0}\rotatebox{0}{\makebox(0,0)[cb]{\smash{$RMS_{ref} $}}}}
   \put(0.280,0.200){\color[rgb]{0,0,0}\rotatebox{0}{\makebox(0,0)[cb]{\smash{$\sqrt{P}$}}}}
   \put(0.990,0.500){\color[rgb]{0,0,0}\rotatebox{90}{\makebox(0,0)[cb]{\smash{$RMS_{ref}  \; \longrightarrow$}}}}
   \put(0.040,0.500){\color[rgb]{0,0,0}\rotatebox{90}{\makebox(0,0)[cb]{\smash{$\sqrt{P} \; \longrightarrow$}}}}
   \put(0.055,0.070){\color[rgb]{0,0,0}\rotatebox{0}{\makebox(0,0)[rb]{\smash{$10^{-\frac{5}{2}}$}}}}
   \put(0.055,0.910){\color[rgb]{0,0,0}\rotatebox{0}{\makebox(0,0)[rb]{\smash{$10^{\frac{1}{2}}$}}}}
   \put(0.940,0.070){\color[rgb]{0,0,0}\rotatebox{0}{\makebox(0,0)[lb]{\smash{$10^{-3}$}}}}
   \put(0.940,0.910){\color[rgb]{0,0,0}\rotatebox{0}{\makebox(0,0)[lb]{\smash{$10^{-\frac{1}{2}}$}}}}
   \put(0.080,0.022){\color[rgb]{0,0,0}\rotatebox{0}{\makebox(0,0)[cb]{\smash{$0$}}}}
   \put(0.920,0.022){\color[rgb]{0,0,0}\rotatebox{0}{\makebox(0,0)[cb]{\smash{$56$}}}}
   \put(0.500,0.020){\color[rgb]{0,0,0}\rotatebox{0}{\makebox(0,0)[cb]{\smash{computation time (seconds)}}}}
   \end{picture}
   \label{tl1000n3}
}
\; \; \; \;
\subfigure[$L = 11$, $\Omega = 4$, $\mathrm{Re} = 1000$]{
   \begin{picture}(1,1)
   \put(0,0){\includegraphics[width=\unitlength]{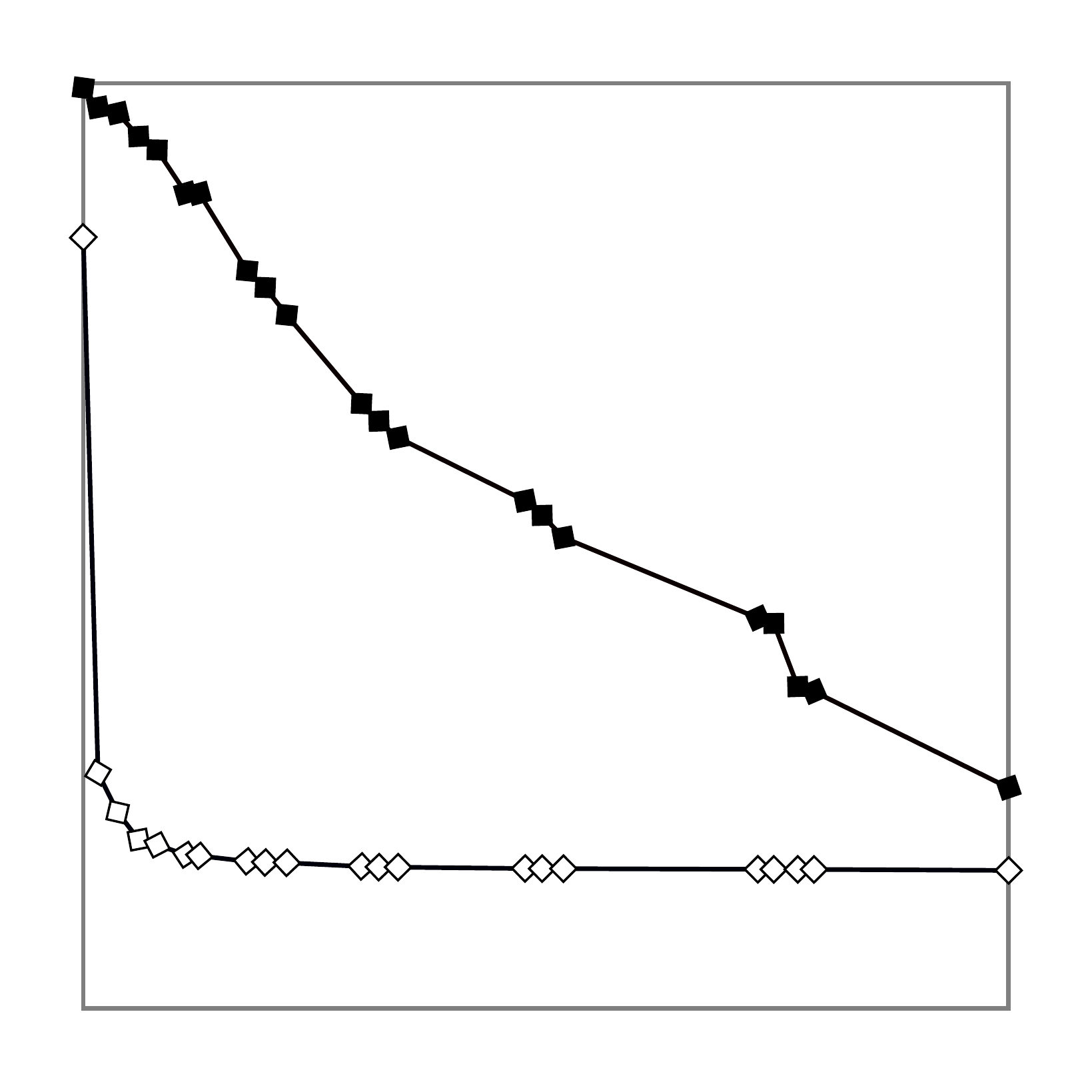}}
   \put(0.620,0.540){\color[rgb]{0,0,0}\rotatebox{0}{\makebox(0,0)[cb]{\smash{$RMS_{ref}$}}}}
   \put(0.380,0.250){\color[rgb]{0,0,0}\rotatebox{0}{\makebox(0,0)[cb]{\smash{$\sqrt{P}$}}}}
   \put(0.990,0.500){\color[rgb]{0,0,0}\rotatebox{90}{\makebox(0,0)[cb]{\smash{$RMS_{ref}  \; \longrightarrow$}}}}
   \put(0.040,0.500){\color[rgb]{0,0,0}\rotatebox{90}{\makebox(0,0)[cb]{\smash{$\sqrt{P} \; \longrightarrow$}}}}
   \put(0.055,0.070){\color[rgb]{0,0,0}\rotatebox{0}{\makebox(0,0)[rb]{\smash{$10^{-\frac{5}{2}}$}}}}
   \put(0.055,0.910){\color[rgb]{0,0,0}\rotatebox{0}{\makebox(0,0)[rb]{\smash{$10^{\frac{1}{2}}$}}}}
   \put(0.940,0.070){\color[rgb]{0,0,0}\rotatebox{0}{\makebox(0,0)[lb]{\smash{$10^{-3}$}}}}
   \put(0.940,0.910){\color[rgb]{0,0,0}\rotatebox{0}{\makebox(0,0)[lb]{\smash{$10^{-1}$}}}}
   \put(0.080,0.022){\color[rgb]{0,0,0}\rotatebox{0}{\makebox(0,0)[cb]{\smash{$0$}}}}
   \put(0.920,0.022){\color[rgb]{0,0,0}\rotatebox{0}{\makebox(0,0)[cb]{\smash{$38$}}}}
   \put(0.500,0.020){\color[rgb]{0,0,0}\rotatebox{0}{\makebox(0,0)[cb]{\smash{computation time (seconds)}}}}
   \end{picture}
   \label{tl1000n4}
}

\subfigure[$L = 200$, $\Omega = 2$, $\mathrm{Re} = 5000$]{
   \begin{picture}(1,1)
   \put(0,0){\includegraphics[width=\unitlength]{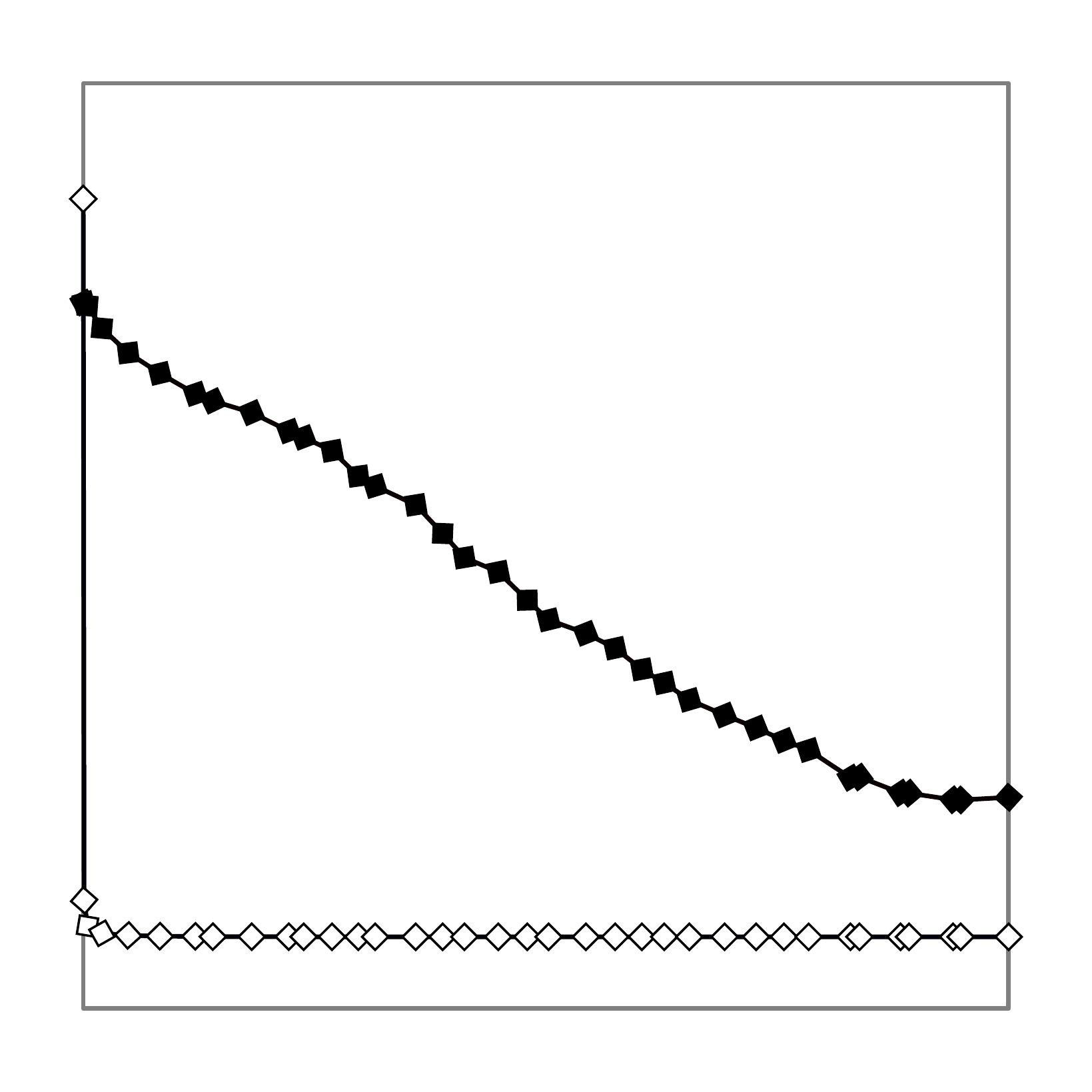}}
   \put(0.580,0.490){\color[rgb]{0,0,0}\rotatebox{0}{\makebox(0,0)[cb]{\smash{$RMS_{ref} $}}}}
   \put(0.280,0.190){\color[rgb]{0,0,0}\rotatebox{0}{\makebox(0,0)[cb]{\smash{$\sqrt{P}$}}}}
   \put(0.990,0.500){\color[rgb]{0,0,0}\rotatebox{90}{\makebox(0,0)[cb]{\smash{$RMS_{ref}  \; \longrightarrow$}}}}
   \put(0.040,0.500){\color[rgb]{0,0,0}\rotatebox{90}{\makebox(0,0)[cb]{\smash{$\sqrt{P} \; \longrightarrow$}}}}
   \put(0.055,0.070){\color[rgb]{0,0,0}\rotatebox{0}{\makebox(0,0)[rb]{\smash{$10^{-\frac{7}{2}}$}}}}
   \put(0.055,0.910){\color[rgb]{0,0,0}\rotatebox{0}{\makebox(0,0)[rb]{\smash{$10^{\frac{1}{2}}$}}}}
   \put(0.940,0.070){\color[rgb]{0,0,0}\rotatebox{0}{\makebox(0,0)[lb]{\smash{$10^{-\frac{5}{2}}$}}}}
   \put(0.940,0.910){\color[rgb]{0,0,0}\rotatebox{0}{\makebox(0,0)[lb]{\smash{$10^{-1}$}}}}
   \put(0.080,0.022){\color[rgb]{0,0,0}\rotatebox{0}{\makebox(0,0)[cb]{\smash{$0$}}}}
   \put(0.920,0.022){\color[rgb]{0,0,0}\rotatebox{0}{\makebox(0,0)[cb]{\smash{$17$}}}}
   \put(0.500,0.020){\color[rgb]{0,0,0}\rotatebox{0}{\makebox(0,0)[cb]{\smash{computation time (hours)}}}}
   \end{picture}
   \label{tl5000n2}
}
\; \; \; \;
\subfigure[$L = 40$, $\Omega = 5$, $\mathrm{Re} = 5000$]{
   \begin{picture}(1,1)
   \put(0,0){\includegraphics[width=\unitlength]{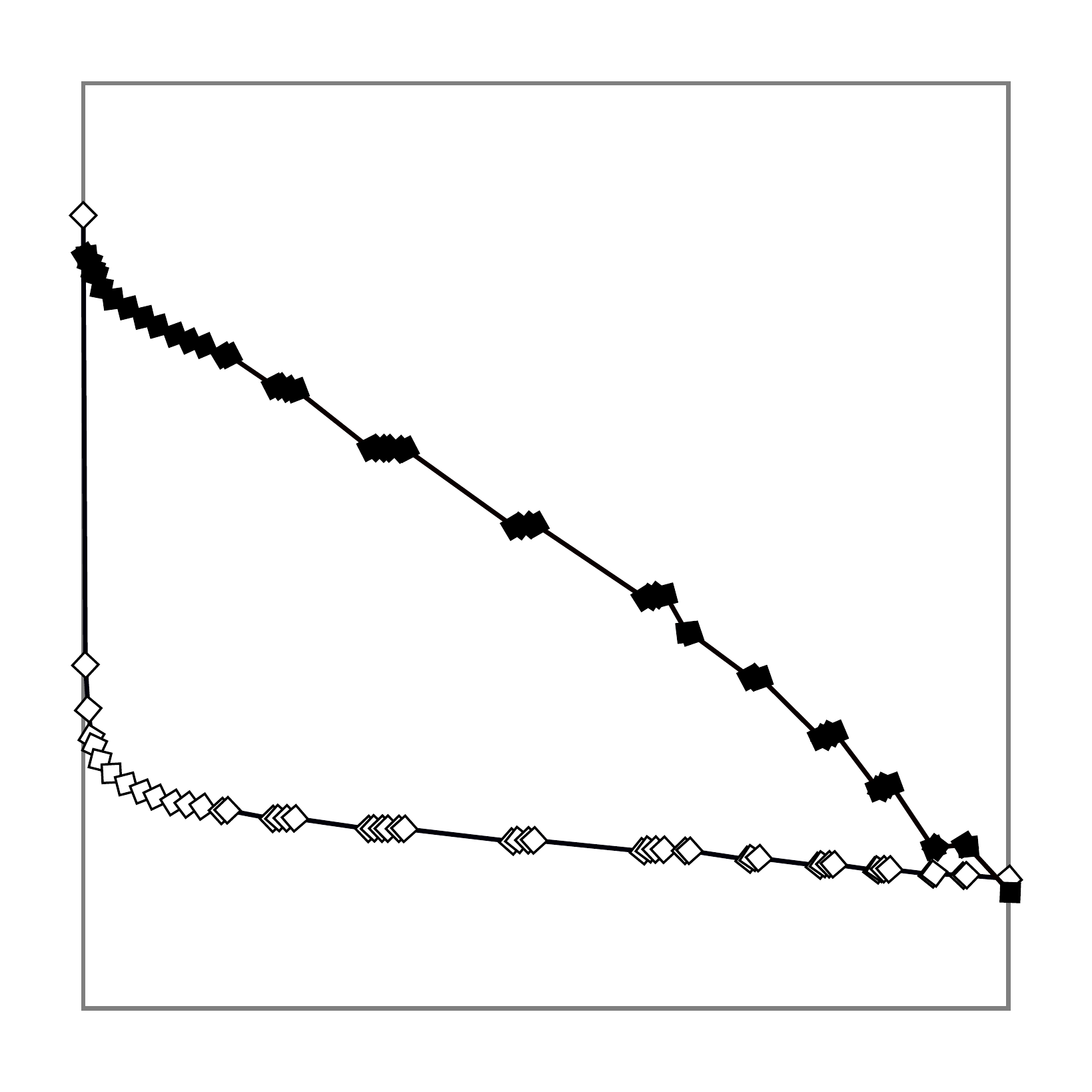}}
   \put(0.600,0.530){\color[rgb]{0,0,0}\rotatebox{0}{\makebox(0,0)[cb]{\smash{$RMS_{ref}$}}}}
   \put(0.390,0.270){\color[rgb]{0,0,0}\rotatebox{0}{\makebox(0,0)[cb]{\smash{$\sqrt{P}$}}}}
   \put(0.990,0.500){\color[rgb]{0,0,0}\rotatebox{90}{\makebox(0,0)[cb]{\smash{$RMS_{ref}  \; \longrightarrow$}}}}
   \put(0.040,0.500){\color[rgb]{0,0,0}\rotatebox{90}{\makebox(0,0)[cb]{\smash{$\sqrt{P} \; \longrightarrow$}}}}
   \put(0.055,0.070){\color[rgb]{0,0,0}\rotatebox{0}{\makebox(0,0)[rb]{\smash{$10^{-3}$}}}}
   \put(0.055,0.910){\color[rgb]{0,0,0}\rotatebox{0}{\makebox(0,0)[rb]{\smash{$10^{\frac{1}{2}}$}}}}
   \put(0.940,0.070){\color[rgb]{0,0,0}\rotatebox{0}{\makebox(0,0)[lb]{\smash{$10^{-3}$}}}}
   \put(0.940,0.910){\color[rgb]{0,0,0}\rotatebox{0}{\makebox(0,0)[lb]{\smash{$10^{-1}$}}}}
   \put(0.080,0.022){\color[rgb]{0,0,0}\rotatebox{0}{\makebox(0,0)[cb]{\smash{$0$}}}}
   \put(0.920,0.022){\color[rgb]{0,0,0}\rotatebox{0}{\makebox(0,0)[cb]{\smash{$193$}}}}
   \put(0.500,0.020){\color[rgb]{0,0,0}\rotatebox{0}{\makebox(0,0)[cb]{\smash{computation time (minutes)}}}}
   \end{picture}
   \label{tl5000n5}
}
\caption{These figures show the error (empty squares) and the deviance (solid squares) from a reference solution \citep{Erturk05} on a logarithmic scale for four separate computations, run through several iterations until comparable accuracy was reached. The error, $\sqrt{P}$, (plotted with empty squares) is the root--mean--square grid-cell error (numerically integrated over the entire grid). The quantity, $P$, is given in Eq.(\ref{coefInt}) as a function of the parameters $\theta_u$,$\theta_v$ and $\theta_p$ which are determined as explained in Subsection \ref{nonlincg}. The quantity, $RMS_{ref}$, (plotted with solid squares) is the root--mean--square deviance of the computed solution as compared with the figures given by \citet{Erturk05} for the $x$-component of the velocity along a vertical line through the geometric center of the cavity.  
Subfigures \ref{tl1000n3} and \ref{tl1000n4} shows these quantities over 25 iterations using $L = 20, \Omega = 3$ and $L = 11, \Omega = 4$, respectively, and with Reynolds number, $\mathrm{Re} = 1000$, where an approximate solution for Reynolds number, $\mathrm{Re} = 400$, was used to initialize the flow components. Subfigures \ref{tl5000n2} and \ref{tl5000n5} shows these quantities over 30 and 50 iterations using $L = 200, \Omega = 2$ and $L = 40, \Omega = 5$, respectively, with Reynolds number, $\mathrm{Re} = 5000$, where an approximate solution for Reynolds number, $\mathrm{Re} = 2500$, was used to initialize the flow components.}
\label{timelog}
\end{figure}

Figure \ref{timelog} shows examples of accuracy versus computation time for different Reynolds numbers, grid resolutions and order of continuity. The linear conjugate gradient iteration (see Subsection \ref{linearCG}) accounted for most of the computation time (typically about 95 \%). For higher Reynolds numbers, a higher grid resolution was required to achieve convergence. For Reynolds number, $Re > 5000$, computations were only carried out with $\Omega = 5$ since the required grid resolution for lower values of $\Omega$ made computations on the current system too time consuming. For the highest Reynolds number, $Re = 40000$, the computation was run for approximately $72$ hours with $L = 135$ and $\Omega = 5$. 

The error, $\sqrt{P}$, plotted with empty squares in Figure \ref{timelog}, shows a rapid initial convergence followed by a much slower rate of convergence. It is clear, however, when comparing with the reference figures from \citet{Erturk05}, plotted with solid squares in Figure \ref{timelog}, that the computed solutions still undergo changes during the final iterations. This may be explained by the fact that, while the quantity $\sqrt{P}$ only measures how well the solution conforms to the given differential equations (independently) at each point in the computational domain, the quantity $RMS_{ref}$ depends on the value of the solution at specific points which may be affected by the accumulation of small errors elsewhere in the computational domain. Additionally, some areas of the computational domain (e.g. near the lower corners where the velocity is discontinuous) may suffer from large errors compared to the rest of the grid, dominating the value of the quantity $\sqrt{P}$ in the later stages of the iteration. The residual of the solutions computed by the references \citep{Erturk05,Wahba12} converge to a smaller factor than the error of the current solutions. However, the error \emph{between} grid points is not taken into consideration by \citet{Erturk05,Wahba12}, whereas in the current work the error is computed over a large number of sub-grid sample points.

It is clear that both an increase of the grid resolution, $L$, and the order of continuity, $\Omega-1$, increases the size of the linear system, $\mathbf{\Pi}$ (see Eq.(\ref{gridLinSys})), and thus the required computation time and the required amount of memory. It should also be noted, however, that increasing the order of continuity (i.e. increasing $\Omega$) also increases the number of nonzero sub/super-diagonals of the linear system, $\mathbf{\Pi}$, which also increases memory requirements and computation time. Despite this disadvantage for higher orders of continuity, it was found that using higher values of $\Omega$ was more efficient at computing solutions for high Reynolds numbers ($\mathrm{Re} \gtrapprox 2500$), yielding solutions of acceptable accuracy for lower values of the grid resolution, $L$. The value of $\Omega$ is limited by the floating point errors, as shown in Subsection \ref{basisChoice}. For this reason, a maximum value of $5$ was chosen (i.e. $\Omega \leq 5$ in all computations), corresponding to a spatial 4'th order of continuity ($C^4$) and a polynomial accuracy of order 9.

\section{Conclusion and Outlook}
An increase in spatial order (polynomial degree) of the grid ($p$-refinement) has advantages compared to increasing the grid resolution ($h$-refinement) in some cases when using the current method, as shown by the high $\mathrm{Re}$ solutions for the lid-driven cavity. These solutions, computed on an ordinary desktop computer, are among the highest Reynolds numbers at which steady state solutions for the lid-driven cavity have been published, even though obvious optimizations (e.g. mesh grading or parallel computation) were not used.

Unlike pseudo-time finite differencing approaches, the current method for arriving at a steady state solution does not yield periodic solutions as artifacts. Instability may appears if the grid resolution is insufficient, but it is chaotic, and does not resemble a periodic flow configuration. 

It is clear from physical evidence that, for the high Reynolds numbers ($\mathrm{Re} \gtrapprox 5000$), the presented steady state solutions do not correspond to a physical three dimensional flow. It is, however, interesting to note that small perturbations, which are thought to initiate turbulence in a real flow, may be mimicked by numerical inaccuracies and potentially initiate turbulence or periodic behavior in simulations. The large differences in reported Reynolds number at which steady state solutions have been obtained for the lid-driven cavity in two dimensions may be explained by the different nature and magnitude of these inaccuracies. If a periodic behavior, observed when solving the two dimensional system, was exclusively due to the mathematical qualities of the of the system (i.e. due to Poincar\'{e}--Andronov--Hopf bifurcation), it is reasonable to assume that this behavior would have occurred at similar Reynolds numbers even though different numerical schemes were used.

The grids with the highest order of continuity, $\Omega = 5$ (equivalent to a polynomial degree of $9$, see Subsection \ref{ordCont}-\ref{basisFuncAndDegree}), were the most efficient for computing steady state solutions for high Reynolds number flows, but the numerical accuracy imposed limitations on further increase of the order of continuity. An improvement, for example by using increased floating point precision or by finding basis functions with better numerical properties, is clearly possible.

It is clear from the mathematical framework (see Section \ref{framework}) that the current method can be generalized to higher dimensions. Further, linear terms (e.g. time derivative, for unsteady flows) may also be added to the governing equations in matrix form (see Subsection \ref{weakform}) without fundamentally changing the properties of the method. With the current method, and other finite-element based methods, one obtains coupled sets of equations depending on information in a grid. The computational cost required to solve these systems tend to grow exponentially with the number of grid points. However, the computational cost of the numerical integration, which defines the equation set for the current method (see Subsection \ref{subCell}) grows linearly with the number of grid points. This is an advantage because, instead of adapting the grid to complex geometry or to different fluid phases, with the current method it is possible to select different governing equations independently at different sample points. One can also increase the density of sample points in some areas if necessary (assuming appropriate weighting is applied). Interaction with objects smaller than the grid scale can thus be incorporated. An interface between immiscible fluid phases can be incorporated in the same way. The latter will be demonstrated with three-dimensional unsteady flow in a forthcoming paper.

\section*{Acknowledgments} 
This work was supported by BKK Production and The Research Council of Norway under The Industrial Ph.D Scheme. 

Alex Hoffmann\footnote{Professor, University of Bergen, MAE}, Jan Vaagen\footnote{Professor, University of Bergen, MAE}, Laszlo Csernai\footnote{Professor, University of Bergen, MAE} and Arne Sm\aa brekke\footnote{Department Manager, BKK Production} are acknowledged for productive discussions and valuable suggestions.


\bibliographystyle{plainnat}
\bibliography{thebib}

\end{document}